\documentclass[twocolumn,english]{IEEEtran}
\usepackage[T1]{fontenc}
\usepackage{babel}
\usepackage{prettyref}
\usepackage{amsthm}
\usepackage{amsmath}
\usepackage{amssymb}
\usepackage{graphicx}
\usepackage[unicode=true,
 bookmarks=true,bookmarksnumbered=true,bookmarksopen=true,bookmarksopenlevel=1,
 breaklinks=false,pdfborder={0 0 0},backref=false,colorlinks=false]
 {hyperref}
\hypersetup{pdftitle={Your Title},
 pdfauthor={Your Name},
 pdfpagelayout=OneColumn, pdfnewwindow=true, pdfstartview=XYZ, plainpages=false}
\usepackage{breakurl}

\makeatletter

\providecommand{\tabularnewline}{\\}

\theoremstyle{plain}
\newtheorem{thm}{\protect\theoremname}
\theoremstyle{definition}
\newtheorem{defn}[thm]{\protect\definitionname}

\ifCLASSOPTIONcompsoc
\usepackage[caption=false,font=normalsize,labelfont=sf,textfont=sf]{subfig}
\else
\usepackage[caption=false,font=footnotesize]{subfig}
\fi
\usepackage{cite}
\usepackage{amsthm}
\newtheorem{theorem}{\textbf{Theorem}}

\newtheorem{lemma}{\textbf{Lemma}}

\@ifundefined{showcaptionsetup}{}{%
 \PassOptionsToPackage{caption=false}{subfig}}
\usepackage{subfig}
\makeatother

\providecommand{\definitionname}{Definition}
\providecommand{\theoremname}{Theorem}

\begin{document}

\title{D2D Enhanced Heterogeneous Cellular Networks with Dynamic TDD}

\author{Hongguang~Sun, ~Matthias Wildemeersch,~\IEEEmembership{Member,~IEEE,}~\\
Min Sheng,~\IEEEmembership{Member,~IEEE,}$\:$Tony Q. S. Quek, ~\IEEEmembership{Senior Member,~IEEE}
\thanks{H. Sun and M. Sheng (corresponding author) are with the State Key
Laboratory of Integrated Service Networks, Institute of Information
Science, Xidian University, Xi'an, Shaanxi, 710071, China. (email:
hgsun@xidian.edu.cn, msheng@mail.xidian.edu.cn).

Matthias Wildemeersch is with the Singapore University of Technology
and Design, Singapore. (e-mail: m.w.wildemeersch@ieee.org).

T. Q. S. Quek is with the Singapore University of Technology and Design
and the Institute for Infocomm Research, Singapore. (e-mail: tonyquek@sutd.edu.sg).%
}}
\maketitle
\begin{abstract}
Over the last decade, the growing amount of UL and DL mobile data
traffic has been characterized by substantial asymmetry and time variations.
Dynamic time-division duplex (TDD) has the capability to accommodate
to the traffic asymmetry by adapting the UL/DL configuration to the
current traffic demands. In this work, we study a two-tier heterogeneous
cellular network (HCN) where the macro tier and small cell tier operate
according to a dynamic TDD scheme on orthogonal frequency bands. To
offload the network infrastructure, mobile users in proximity can
engage in D2D communications, whose activity is determined by a carrier
sensing multiple access (CSMA) scheme to protect the ongoing infrastructure-based
and D2D transmissions. We present an analytical framework to evaluate
the network performance in terms of \emph{load-aware} coverage probability
and network throughput. The proposed framework allows to quantify
the effect on the coverage probability of the most important TDD system
parameters, such as the UL/DL configuration, the base station density,
and the bias factor. In addition, we evaluate how the bandwidth partition
and the D2D network access scheme affect the total network throughput.
Through the study of the tradeoff between coverage probability and
D2D user activity, we provide guidelines for the optimal design of
D2D network access. \end{abstract}


\section{Introduction}

As social applications in current data-centric networks continue to
increase, mobile operators need to address the exponential growth
of data traffic. Deploying diverse low-power small cell\emph{ }access
points (SAPs) to complement the conventional macrocell network has
proven as a cost-effective means to increase the network capacity
and enhance coverage \cite{SCND,TTHW,OTCOD,CSCN}. Another technique
to address the explosion of mobile data traffic is device-to-device
(D2D) communications \cite{DDCI,GRAM_Song}. With this technology,
mobile users in proximity can establish a direct link and bypass the
base stations, thereby offloading the network infrastructure and providing
increased spectral efficiency \cite{SSFD,AMMS,ROID2D,ERAF_Song,CGFRA,OTCR}.
With tools from stochastic geometry \cite{SGAI}, tractable analytical
frameworks were developed for the design of D2D spectrum sharing in
frequency-division duplex (FDD) cellular networks \cite{SSFD,AMMS,ROID2D,OTCR}.
Game theoretic models were applied to study the resource allocation
for D2D communication in \cite{GRAM_Song}, \cite{ERAF_Song}, \cite{CGFRA}.
Specifically, Xu et al. \cite{ERAF_Song} proposed a reverse iterative
combinatorial auction based approach to efficiently assign the downlink
(DL) cellular resource to D2D users. The uplink (UL) resource allocation
issue between D2D and cellular users was investigated in \cite{CGFRA},
in which a coalition formation game model was proposed to maximize
the system sum rate. Aside from the surge in data traffic, Internet
services and video applications also lead to asymmetry and dynamic
variations in the UL and DL traffic load. Time-division duplex (TDD)
systems \cite{LTUL} have the capability to manage the UL/DL traffic
asymmetry by adjusting the fraction of time dedicated to UL and DL
transmissions, which we refer to as the UL/DL configuration, to the
current traffic conditions. To accommodate the instantaneous traffic
load among different cells, dynamic TDD with variable UL/DL configuration
is under consideration and allows to make better use of the resources
\cite{DUDC,DUDO}.

In a two-tier HCN operating with universal frequency reuse, the major
challenge is the cross-tier and co-tier interference. In addition,
extra interference is imposed to the cellular transmissions if underlaid
D2D transmissions are admitted. In a network operating with dynamic
TDD, interference conditions can be severe and strong DL-to-UL (base
station-to-base station) interference may lead to an unacceptable
performance in the UL transmissions \cite{DTAF}. Considering the
base stations and mobile users distribute as poisson point processes
(PPPs), Yu et al. \cite{DTSI} derived the distributions of DL and
UL SINR at an arbitrary mobile user and base station in a single tier
dynamic TDD small cell network. However, the effect of important parameters
such as UL/DL configuration and base station density on the network
performance was not analyzed. Considering a two-tier HCN, a cognitive
hybrid division duplex (CHDD) scheme was proposed in \cite{CHDD},
where the macrocells operate with FDD, and small cells operate dynamic
TDD on both FDD bands. Without interference management, \cite{CHDD}
demonstrates that universal frequency reuse leads to a significant
deterioration in the UL signal quality. To alleviate the cross-tier
interference in a two-tier HCN, variable interference management schemes
have been proposed, such as power control \cite{PCIT}, interference
cancellation \cite{SICIH}, and spectrum allocation \cite{SAIT}.
For spectrum allocation, a distributed disjoint subchannel allocation
policy is sensible especially in dense small cell networks \cite{SAIT}.
To address the co-tier interference, medium access control (MAC) is
an effective and widely used technique in distributed ad hoc/sensor
networks \cite{ASGM,MIIH,AMHC,IAOI}. Carrier sensing multiple access
(CSMA) is a popular MAC protocol where the positions of simultaneously
transmitting nodes can be modeled by a \emph{Matern Hard-core Process
(MHP)} \cite{ASGM,MIIH,AMHC}. In an MHP, each node respects a minimum
exclusion distance with respect to each other so as to control the
mutual interference. Carrier sensing is also employed in cognitive
radio networks to limit the interference inflicted on primary users
(PUs). In \cite{IAOI}, secondary users (SUs) are modeled as a\emph{
Poisson Hole Process (PHP)}, such that only SUs located outside the
exclusion region of PUs can transmit.

Despite the fact that both the merit of dynamic TDD networks \cite{DUDC,DUDO},
and the benefit of D2D communications in FDD networks \cite{SSFD,AMMS,ROID2D}
have been widely discussed in literature, a unifying framework for
D2D enhanced TDD networks is still missing. In this work, we consider
a D2D enhanced two-tier HCN operating with dynamic TDD where macrocells
and small cells operate on two orthogonal frequency bands to eliminate
the cross-tier interference. D2D users share the same bandwidth with
the small cell tier and control their interference by means of a CSMA
scheme. Furthermore, prior literature usually considers a fully-loaded
model where every Voronoi cell has mobile users to connect. However,
due to the small coverage of SAPs, the fully-loaded model may significantly
overstate the network interference from small cells, leading to a
pessimistic estimation on the coverage probability. In this work,
we consider a \emph{load-aware} model, where the empty cells are considered
and the density of \emph{active} cells is derived. Our main contributions
can be listed as follows:
\begin{itemize}
\item We propose a simple PPP model for the \emph{active} D2D transmitters
based on the combined effect of a PHP and an MHP process, and we illustrate
the validity of the PPP approximation by means of extensive simulations.
\item We define an association policy that decouples the cell associations
in UL and DL, and present an analytical framework that describes the
\emph{load-aware} coverage probability and network throughput as a
function of all relevant system parameters. Although the effect of
base station density and bias factor on the coverage probability is
well understood in FDD networks \cite{HCNW}, the effect of these
system parameters in dynamic TDD networks is still unclear.
\item We evaluate the effect of D2D network access scheme on network performance,
and quantify its advantage over the random access scheme ALOHA.
\item We study the tradeoff between coverage probability and D2D user activity.
From the perspective of total network throughput, we provide guidelines
for the optimal design of the network access scheme in D2D enhanced
TDD networks.
\end{itemize}
The rest of the paper is organized as follows. In Section II, the
system model is presented. In Section III, the load-aware coverage
probability and network throughput are derived and the validity of
the analytical framework is demonstrated by means of the numerical
simulations. In Section IV, the impact of key parameters on the network
coverage probability is evaluated and operating guidelines for the
practical network design are provided. In Section V, the effect of
bandwidth partition between the tiers and the impact of D2D network
access control on the network performance are evaluated. Conclusions
are given in Section VI.

\section{System Model}

\subsection{Network Model}

\begin{figure}[t]
\centering\includegraphics[scale=0.5]{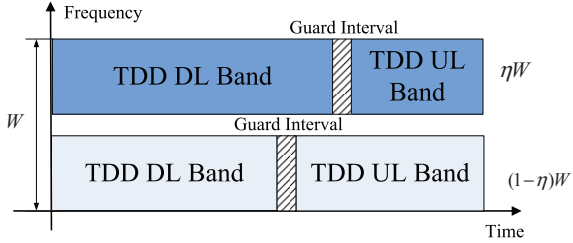}\protect\caption{\label{fig:Bandwidth Partition}Dynamic TDD scheme in two-tier heterogeneous
cellular networks.}
\end{figure}

We consider a two-tier HCN which consists of a first tier of macro
base stations (MBSs) distributed according to a homogeneous PPP $\Phi_{\mathtt{m}}$
with density $\lambda_{\mathtt{m}}$, overlaid with a network of SAPs
distributed according to a PPP $\Phi_{\mathtt{s}}$ with density $\lambda_{\mathtt{s}}$.
Mobile users are scattered over $\mathbb{R}^{2}$ according to a PPP
$\Phi_{\mathtt{u}}$ with density $\lambda_{\mathtt{u}}$. A fraction
$\zeta$ of the mobile users have their target receiver within a close
distance and are considered as potential D2D transmitters. As an independent
thinning of $\Phi_{\mathtt{u}}$ with probability $\zeta$, the set
of potential D2D transmitters $\tilde{\Phi}_{\mathtt{d}}=\{T_{i}\}$
forms a PPP with density $\zeta\lambda_{\mathtt{u}}$. We assume that
each potential D2D transmitter has an assigned receiver (not belonging
to $\Phi_{\mathtt{u}}$) at a fixed distance $r_{\mathtt{d}}$ in
a uniformly random direction.%
\footnote{We note that the potential D2D receivers are scattered according to
a PPP with density $\zeta\lambda_{\mathtt{u}}$, where the potential
D2D receivers and $\Phi_{\mathtt{u}}$ are dependent point processes.%
} We consider orthogonal spectrum allocation where the total bandwidth
$W$ is divided into two non-overlapping parts $\eta W$ and $\left(1-\eta\right)W$
that are allocated to the macro tier and small cell tier as depicted
in Fig. \ref{fig:Bandwidth Partition}. The potential D2D users share
the spectrum with the small cell tier, thus leading to coexistence
issues with the small cell users and SAPs. Both the macro tier and
small cell tier operate according to the dynamic TDD scheme where
at each timeslot a cell configures flexibly in DL or UL mode. The
transmission mode selection for macrocells and small cells is modeled
by independent Bernoulli random variables (r.v.'s), such that macrocells
and small cells are configured in DL mode with probability $q_{\mathtt{D},\mathtt{m}}$
and $q_{\mathtt{D},\mathtt{s}}$, respectively, while the corresponding
UL mode probabilities are given by $1-q_{\mathtt{D},\mathtt{m}}$
and $1-q_{\mathtt{D},\mathtt{s}}$. The multiplexing probabilities
$q_{\mathtt{D},\mathtt{m}}$ and $q_{\mathtt{D},\mathtt{s}}$ define
the UL/DL configuration for the macro tier and small cell tier. The
concurrent DL and UL transmissions in neighboring cells may lead to
new types of inter-cell interference, i.e. DL-to-UL and UL-to-DL (user-to-user)
interference. Let $P_{\mathtt{m}}$, $P_{\mathtt{s}}$, $Q_{\mathtt{m}}$
and $Q_{\mathtt{s}}$ denote the transmit power of MBSs, SAPs, mobile
users associated with the macro tier, and mobile users associated
with the small cell tier.%
\footnote{Note that in this work we consider a baseline model that does not
account for UL power control. However, it is possible to extend the
network performance analysis to the case with power control policy
by applying the results developed in \cite{SSFD,AMMS,AMUC,OTCR}. %
} We use $Q_{\mathtt{d}}$ to represent the transmit power of potential
D2D users. To avoid confusion, in the following parts the term \emph{mobile
users} only denotes the mobile users that are expected to communicate
via infrastructures, while the term \emph{potential D2D users }refers
to mobile users which attempt to communicate with each other by employing
D2D technology.

We consider a load-aware resource allocation model where each base
station always has data to transmit if it has a mobile user within
its coverage. We adopt orthogonal multiple access, such that within
a cell only a single mobile user can be active at any given timeslot
and subchannel. If several mobile users connect to the same base station,
the base station will randomly choose one mobile user to serve. To
control the interference inflicted by D2D transmissions on small cell
transmissions, we provide medium access control by means of a CSMA
scheme. The channel model consists of path loss and flat-fading. The
fading power from a transmitter located at point $x$ to the typical
receiver located at the origin is denoted by $h_{ox}$ and is assumed
to be an independent and identically distributed (i.i.d.) exponential
r.v., $h\sim\textrm{exp(1)}$, which corresponds to Rayleigh fading.
The path loss function is given by $g\left(\Vert x\Vert\right)=\Vert x\Vert^{-\alpha}$,
with $\alpha>2$ the path loss exponent. Due to the slight impact
of thermal noise in current heterogeneous networks, we consider the
interference-limited regime, and ignore the thermal noise \cite{ATAT}.

\subsection{Cell Association}

At each timeslot, a mobile user acts as a transmitter or receiver
with probability $\mu$ and $1-\mu$, respectively. Assuming open
access, the association of a mobile user to a given tier is based
on the maximum biased received signal power averaged over fading.
The bias factor in this association policy is used to balance the
traffic load among different tiers. In this paper, we consider a decoupled
DL and UL association model as follows.

\subsubsection{Downlink Association Policy}

A typical receiver is associated with the nearest base station in
DL mode of tier\emph{ $i$ }if
\begin{equation}
i=\arg\underset{k\in\{\mathtt{m},\,\mathtt{s}\}}{\max}P_{k}B_{\mathtt{D},k}D_{\mathtt{D},k}^{-\alpha},\label{eq:Ass_tierI_DL}
\end{equation}
where $B_{\mathtt{D},k}$ is the DL bias factor of tier $k$, and
$D_{\mathtt{D},k}$ denotes the distance from the typical receiver
to the nearest base station of $\Phi_{k}$ operating in DL mode with
thinned density $q_{\mathtt{D},k}\lambda_{k}$.

\subsubsection{Uplink Association Policy}

A typical transmitter is associated with the nearest base station
in UL mode of tier $i$ if
\begin{equation}
i=\arg\underset{k\in\{\mathtt{m},\,\mathtt{s}\}}{\max}Q_{k}B_{\mathtt{U},k}D_{\mathtt{U},k}^{-\alpha},\label{eq:Ass_tierI_UL}
\end{equation}
where $B_{\mathtt{U},k}$ is the UL bias factor of tier $k$, $D_{\mathtt{U},k}$
denotes the distance from the typical transmitter to the nearest base
station of $\Phi_{k}$ operating in UL mode with thinned density $\left(1-q_{\mathtt{D},k}\right)\lambda_{k}$.

For notational brevity, we define the normalized parameters of tier
$k$ conditioned on the serving tier $i$.
\[
\hat{\lambda}_{k}^{(i)}\triangleq\frac{\lambda_{k}}{\lambda_{i}},\:\hat{q}_{\mathtt{D},k}^{(i)}\triangleq\frac{q_{\mathtt{D},k}}{q_{\mathtt{D},i}},\:\hat{q}_{\mathtt{U},k}^{(i)}\triangleq\frac{1-q_{\mathtt{D},k}}{1-q_{\mathtt{D},i}},\:
\]
\begin{equation}
\hat{P}_{k}^{(i)}\triangleq\frac{P_{k}}{P_{i}},\:\hat{Q}_{k}^{(i)}\triangleq\frac{Q_{k}}{Q_{i}},\:\hat{B}_{\mathtt{D},k}^{(i)}\triangleq\frac{B_{\mathtt{D},k}}{B_{\mathtt{D},i}},\:\hat{B}_{\mathtt{U},k}^{(i)}\triangleq\frac{B_{\mathtt{U},k}}{B_{\mathtt{U},i}}.
\end{equation}

Using the association rules defined in \prettyref{eq:Ass_tierI_DL}
and \prettyref{eq:Ass_tierI_UL}, the set of base stations form different
multiplicatively weighted Voronoi tessellations of the two dimensional
plane in DL and UL.
\begin{defn}
The downlink and uplink association regions of an MBS or a SAP located
at point $x$ are given by \prettyref{eq:(4)} and \prettyref{eq:(5)},
respectively \cite{OIHN}
\begin{eqnarray}
\mathcal{C}_{\mathtt{D},x} & = & \Bigl\{ y\in\mathbb{R}^{2}|\Vert y-x\Vert\leq\bigl(\hat{P}_{k}^{(i)}\hat{B}_{\mathtt{D},k}^{(i)}\bigr)^{-\frac{1}{\alpha}}\nonumber \\
 &  & \times\Vert y-X_{\mathtt{D},k}^{*}\left(y\right)\Vert,\forall k\in\{\mathtt{m},\,\mathtt{s}\}\Bigr\},\label{eq:(4)}
\end{eqnarray}
\begin{eqnarray}
\mathcal{C}_{\mathtt{U},x} & = & \Bigl\{ y\in\mathbb{R}^{2}|\Vert y-x\Vert\leq\Bigl(\hat{Q}_{k}^{(i)}\hat{B}_{\mathtt{U},k}^{(i)}\Bigr){}^{-\frac{1}{\alpha}}\nonumber \\
 &  & \times\Vert y-X_{\mathtt{U},k}^{*}\left(y\right)\Vert,\forall k\in\{\mathtt{m},\,\mathtt{s}\}\Bigr\},\label{eq:(5)}
\end{eqnarray}
where $X_{\mathtt{D},k}^{*}\left(y\right)$ and $X_{\mathtt{U},k}^{*}\left(y\right)$
denote the corresponding distances from $y$ to the nearest DL base
station and to the UL base station of tier \emph{k}.
\end{defn}
As a consequence, a mobile user may associate with different base
stations for DL and UL traffic. Let\textbf{ $\mathcal{N}_{\mathtt{D},i}$}
and\textbf{ $\mathcal{N}_{\mathtt{U},i}$} denote the DL load and
UL load, defined as the number of mobile users served by a base station
of tier $i$ operating in DL and UL.

\subsection{CSMA model of potential D2D users}

At the start of a timeslot, each potential D2D transmitter senses
the active small cell transmissions which originate from SAPs in DL
mode and transmitting mobile users associated with the small cell
tier. Assuming channel reciprocity, the potential D2D transmitter
predicts the \emph{would-be }interference it may impose on the small
cell transmitters and refrains from transmitting if the interference
exceeds the protection threshold $\rho_{\mathtt{s}}$. As such, D2D
transmissions respect an exclusion region around each small cell transmitter.
The remaining potential D2D transmitters form a PHP,\emph{}%
\footnote{Note that \cite{IAOI} considers a fixed exclusion distance. In this
work, we account for the channel fading, and thus the exclusion region
of each small cell transmitter is a function of the instantaneous
channel gain.%
} which can be approximated by a PPP \cite{IAOI}. Carrier sensing
is also performed with respect to the remaining potential D2D transmitters,
where the signal power from a nearby D2D transmitter is not allowed
to surpass the contention threshold $\rho_{\mathtt{d}}$. To resolve
the collision among the D2D contenders, we use a back-off scheme.
Specifically, each remaining D2D transmitter independently samples
a random timer $t_{i}\sim\mathcal{U}[0,1]$ and channel access is
granted to the contender with the smallest timer within a contention
region \cite{ASGM}.

Let $U_{i}$ be the retention indicator of the $i$-th potential D2D
transmitter $T_{i}$,%
\footnote{We use $T_{i}$ to indicate the position of the transmitter and the
transmitter itself.%
} which is given by
\begin{eqnarray}
U_{i} & = & \prod_{Y_{j}\in\Phi_{\mathtt{s}}^{\mathtt{D}}}\mathbf{1}_{\bigl(\frac{Q_{\mathtt{d}}h_{ji}}{\Vert T_{i}-Y_{j}\Vert^{\alpha}}<\rho_{\mathtt{s}}\bigr)}\prod_{Z_{l}\in\Phi_{\mathtt{u},\mathtt{s}}^{\mathtt{T}}}\mathbf{1}_{\bigl(\frac{Q_{\mathtt{d}}h_{li}}{\Vert T_{i}-Z_{l}\Vert^{\alpha}}<\rho_{\mathtt{s}}\bigr)}\nonumber \\
 & \times & \prod_{T_{k}\in\Phi_{\mathtt{d}}\backslash T_{i}}\Bigl(\mathbf{1}_{\left(t_{i}\leq t_{k}\right)}+\mathbf{1}_{\left(t_{i}>t_{k}\right)}\mathbf{1}_{\bigl(\frac{Q_{\mathtt{d}}h_{ki}}{\Vert T_{i}-T_{k}\Vert^{\alpha}}<\rho_{\mathtt{d}}\bigr)}\Bigr)\label{eq:(6)}
\end{eqnarray}
where $h_{ji}$ denotes the channel fading from $T_{i}$ to $Y_{j}$,
$\{Y_{j}\}=\Phi_{\mathtt{s}}^{\mathtt{D}}$ denotes the set of active
SAPs in DL, and $\{Z_{l}\}=\Phi_{\mathtt{u},\mathtt{s}}^{\mathtt{T}}$
represents the set of transmitting mobile users associated with the
small cell tier. The first two products in \prettyref{eq:(6)} reflect
that the interference inflicted by a potential D2D transmitter on
an active small cell transmitter should be smaller than $\rho_{\mathtt{s}}$.
The first term inside the last product corresponds to the event where
the timer $t_{i}$ is smaller than $t_{k}$, while the second term
corresponds to the event where the timer $t_{i}$ is larger than $t_{k}$,
yet the interference from $T_{i}$ to $T_{k}$ is smaller than $\rho_{\mathtt{d}}$.

Define $\beta\overset{\triangle}{=}\Pr[U_{i}=1]$ as the retaining
probability of $T_{i}$, which depends on the timer $t_{i}$, the
channel fading and the distance between $T_{i}$ and small cell transmitters.
The set of winning contenders forms a point process similar to the
MHP,%
\footnote{By definition, the MHP originates from a homogeneous PPP with some
density $\lambda$, where each node associates with a random mark.
A node is forbidden to transmit only if there is another node within
a certain exclusion distance with a smaller mark \cite{MIIH}. %
} where any two points respect a minimum exclusion distance determined
by $\rho_{\mathtt{d}}$ and the instantaneous channel gain. It is
known that the aggregate interference experienced by a user of an
MHP can be approximated by the interference resulting from a PPP that
has the same density as the MHP and exists outside the exclusion region
\cite{MIIH}, \cite{AMHC}. In this work, we \textit{\emph{assume}}
that each potential D2D transmitter is retained independently with
the probability $\beta$.%
\footnote{This assumption is an approximation since the retention of D2D transmitters
by means of the CSMA scheme results in a dependent thinning of the
original PPP.%
} As a result, the retained D2D transmitters $\Phi_{\mathtt{d}}$ form
a PPP with density $\beta\zeta\lambda_{\mathtt{u}}$, where the combined
effect of PHP and MHP is captured by $\beta$. Note that the retained
D2D transmitters are the actually\emph{ active} D2D transmitters in
the current timeslot. The potential D2D transmitters that fail to
access channel will keep silent in the current timeslot and continue
to execute the CSMA scheme in the next timeslot.

\section{Performance Analysis}

In this section, we derive the load-aware coverage probability and
network throughput, and we validate the theoretical model by means
of simulations.

\subsection{Association and Load Characterization}

The probability that a typical receiving and transmitting mobile user
is associated with tier $i$ for DL and UL, is given by
\begin{eqnarray}
\mathcal{A}_{\mathtt{D},i} & = & \frac{q_{\mathtt{D},i}\lambda_{i}}{\sum_{k\in\{\mathtt{m},\mathtt{s}\}}G_{\mathtt{D},k}^{(i)}},\:\mathcal{A}_{\mathtt{U},i}=\frac{\left(1-q_{\mathtt{D},i}\right)\lambda_{i}}{\sum_{k\in\{\mathtt{m},\mathtt{s}\}}G_{\mathtt{U},k}^{(i)}},\label{eq:Asso_D_U}
\end{eqnarray}
where $G_{\mathtt{D},k}^{(i)}=q_{\mathtt{D},k}\lambda_{k}\left(\hat{P}_{k}^{(i)}\hat{B}_{\mathtt{D},k}^{(i)}\right)^{\frac{2}{\alpha}},\: G_{\mathtt{U},k}^{(i)}=\left(1-q_{\mathtt{D},k}\right)\lambda_{k}\left(\hat{Q}_{k}^{(i)}\hat{B}_{\mathtt{U},k}^{(i)}\right)^{\frac{2}{\alpha}}.$
The result is a corollary of Lemma 1 in \cite{HCNW} and extends the
DL association policy to the dynamic TDD scheme. For the special case
of $\{q_{\mathtt{D,m}},q_{\mathtt{D,s}}\}=\{0,0\}$, we define $\{\mathcal{A}_{\mathtt{D},\mathtt{m}},\mathcal{A}_{\mathtt{D},\mathtt{s}}\}=\{0,0\}$,
and the network changes into a two-tier UL network. For $\{q_{\mathtt{D,m}},q_{\mathtt{D,s}}\}=\{1,1\}$,
we define $\{\mathcal{A}_{\mathtt{U},\mathtt{m}},\mathcal{A}_{\mathtt{U},\mathtt{s}}\}=\{0,0\}$,
and the network transforms to a two-tier DL network. The association
probabilities defined in \prettyref{eq:Asso_D_U} indicate how the
per tier association probability in a two-tier dynamic TDD network
depends on the \emph{relative} transmit power, bias factor, and base
station density of the corresponding transmission mode. Note that
the base station density affects the per tier association probability
more than transmit power or bias factor.

By considering the traffic load, we derive a more accurate \emph{load-aware}
coverage probability. For each tier \emph{i}, we compute the void
probability of a random base station in DL and UL, i.e. $\mathbb{P}_{\mathit{\mathit{e}}}^{\mathtt{D},i}$
and $\mathbb{P}_{\mathit{\mathit{e}}}^{\mathtt{U},i}$, and compare
it with a threshold value to determine the network traffic load. When
$\mathbb{P}_{\mathit{\mathit{e}}}^{\mathtt{D},i}<10^{-4}$ and $\mathbb{P}_{\mathit{\mathit{e}}}^{\mathtt{U},i}<10^{-4}$,
we say tier \emph{i} is fully-loaded, otherwise, partially-loaded.
Denote $\Phi_{\mathtt{m}}^{\mathtt{D}}\sim\textrm{PPP}(\lambda_{\mathtt{m}}^{\mathtt{D}})$,
$\Phi_{\mathtt{s}}^{\mathtt{D}}\sim\textrm{PPP}(\lambda_{\mathtt{s}}^{\mathtt{D}})$,
$\Phi_{\mathtt{m}}^{\mathtt{U}}\sim\textrm{PPP}(\lambda_{\mathtt{m}}^{\mathtt{U}})$
and $\Phi_{\mathtt{s}}^{\mathtt{U}}\sim\textrm{PPP}(\lambda_{\mathtt{s}}^{\mathtt{U}})$
as the point processes of \emph{active} DL MBSs, DL SAPs, UL MBSs,
and UL SAPs, respectively, with corresponding denstities $\lambda_{\mathtt{m}}^{\mathtt{D}}$,
$\lambda_{\mathtt{s}}^{\mathtt{D}}$, $\lambda_{\mathtt{m}}^{\mathtt{U}}$
and $\lambda_{\mathtt{s}}^{\mathtt{U}}$. In the following lemma,
we derive the void probability of a base station in tier \emph{i},
and we determine the exact density of \emph{active} base stations
in DL and UL.

\begin{lemma}

The probability that a cell of tier $i$ is void for DL and UL is
derived as
\begin{equation}
\mathbb{P}_{\mathit{\mathit{e}}}^{\mathtt{D},i}=\Bigl(1+\frac{\bigl(1-\mu\bigr)\bigl(1-\zeta\bigr)\lambda_{\mathtt{u}}\mathcal{A}_{\mathtt{D},i}}{3.5q_{\mathtt{D},i}\lambda_{i}}\Bigr)^{-3.5},\label{eq:empty_DL}
\end{equation}
\begin{equation}
\mathbb{P}_{\mathit{e}}^{\mathtt{U},i}=\Bigl(1+\frac{\mu\bigl(1-\zeta\bigr)\lambda_{\mathtt{u}}\mathcal{A}_{\mathtt{U},i}}{3.5\left(1-q_{\mathtt{D},i}\right)\lambda_{i}}\Bigr)^{-3.5}.\label{eq:empty_UL}
\end{equation}
Furthermore, the density of active base stations in DL and UL mode
of tier $i$ is given by
\begin{equation}
\lambda_{i}^{\mathtt{D}}=\lambda_{i}q_{\mathtt{D},i}\bigl(1-\mathbb{P}_{e}^{\mathtt{D},i}\bigr)\:\textrm{and}\:\lambda_{i}^{\mathtt{U}}=\lambda_{i}\bigl(1-q_{\mathtt{D},i}\bigr)\bigl(1-\mathbb{P}_{e}^{\mathtt{U},i}\bigr).\label{eq:Density}
\end{equation}

\begin{IEEEproof}
The results can be proved by a minor modification of Lemma 1 in \cite{DCAB}.
Here we give the proof for completeness. The probability density function
(PDF) of the area of a random Voronoi cell is given by $f_{X}(x)=\frac{3.5^{3.5}}{\Gamma(3.5)}x^{2.5}e^{-3.5x}$,
where $X$ denotes the area of a random Voronoi cell normalized by
the value $1/q_{\mathtt{D},i}\lambda_{i}$ in DL and $1/(1-q_{\mathtt{D},i})\lambda_{i}$
in UL mode. Taking the DL mode as an example, the PDF of the DL load
$\mathcal{N}_{\mathtt{D},i}$\textbf{ }is given by
\begin{eqnarray}
 &  & \mathbb{\Pr}[\mathcal{N}_{\mathtt{D},i}=n]\nonumber \\
 & = & \int_{0}^{\infty}\Pr[\mathcal{N}_{\mathtt{D},i}=n|X=x]\cdot f_{X}(x)dx\nonumber \\
 & \overset{(a)}{=} & \int_{0}^{\infty}\frac{1}{n!}\bigl(\frac{\lambda_{\mathtt{u},i}^{\mathtt{R}}x}{q_{\mathtt{D},i}\lambda_{i}}\bigr)^{n}e^{-\frac{\lambda_{\mathtt{u},i}^{\mathtt{R}}x}{q_{\mathtt{D},i}\lambda_{i}}}\cdot f_{X}(x)dx\nonumber \\
 & \overset{(b)}{=} & \frac{3.5^{3.5}}{n!}\frac{\Gamma\left(n+3.5\right)}{\Gamma\left(3.5\right)}\bigl(\frac{\lambda_{\mathtt{u},i}^{\mathtt{R}}}{q_{\mathtt{D},i}\lambda_{i}}\bigr)^{n}\bigl(3.5+\frac{\lambda_{\mathtt{u},i}^{\mathtt{R}}}{q_{\mathtt{D},i}\lambda_{i}}\bigr)^{-\left(n+3.5\right)}\nonumber \\
\label{eq:Load exact}
\end{eqnarray}
where $\lambda_{\mathtt{u},i}^{\mathtt{R}}=\bigl(1-\mu\bigr)\bigl(1-\zeta\bigr)\lambda_{\mathtt{u}}\mathcal{A}_{\mathtt{D},i}$
denotes the density of the receiving mobile users associated with
tier $i$, $q_{\mathtt{D},i}\lambda_{i}$ is the density of total
DL base stations of tier $i$, $\Gamma(\cdot)$ is the gamma function,
which is given by $\Gamma\left(x\right)=\int_{0}^{\infty}t^{x-1}\exp\left(-t\right)dt$,
(a) is due to the definition of Poisson distribution, and (b) takes
the expectation with respect to the area distribution $f_{X}(x)$.
Substituting $n=0$ into \prettyref{eq:Load exact} derives the void
probability $\mathbb{P}_{e}^{\mathtt{D},i}=\Pr[\mathcal{N}_{\mathtt{D},i}=0]=\bigl(1+\frac{\lambda_{\mathtt{u},i}^{\mathtt{R}}}{3.5q_{\mathtt{D},i}\lambda_{i}}\bigr)^{-3.5}$,
which concludes the proof.%
\footnote{To compute the density of active base stations, we consider the void
probability of a random cell, rather than of a typical cell as in
\cite{OIHN}. %
}
\end{IEEEproof}
\end{lemma}

\subsection{Coverage Probability}

With the PPP approximation and the void probability derived in \prettyref{eq:empty_DL},
\prettyref{eq:empty_UL}, we derive the load-aware coverage probability
of tier $i$, $i\in\{\mathtt{m},\:\mathtt{s}\}$ in DL and UL as
\begin{equation}
\mathbb{P}_{i}^{\mathtt{D}}=\Pr[\mathtt{SIR}_{i}^{\mathtt{D}}>\gamma_{i}^{\mathtt{D}}],\;\mathbb{P}_{i}^{\mathtt{U}}=\Pr[\mathtt{SIR}_{i}^{\mathtt{U}}>\gamma_{i}^{\mathtt{U}}],\label{eq:macro coverage}
\end{equation}
where $\gamma_{i}^{\mathtt{D}}$ and $\gamma_{i}^{\mathtt{U}}$ denote
the SIR thresholds of DL and UL transmissions in tier $i$. Similarly,
the coverage probability of a typical D2D receiver is $\mathbb{P}_{\mathtt{d}}=\Pr[\mathtt{SIR}_{\mathtt{d}}>\gamma_{\mathtt{d}}]$
with $\gamma_{\mathtt{d}}$ the SIR threshold of D2D user.

Since we consider open access, the distance between a typical mobile
user and its serving base station of tier $i$ in DL or UL mode, $Y_{\mathtt{D},i}$
or $Y_{\mathtt{U},i}$, is not only influenced by $\Phi_{i}^{\mathtt{D}}$
or $\Phi_{i}^{\mathtt{U}}$, but also by $\Phi_{k}^{\mathtt{D}}$
or $\Phi_{k}^{\mathtt{U}}$, $k\neq i$. The distance distribution
is given by
\begin{equation}
f_{Y_{\mathtt{D},i}}(y)=2\pi\frac{q_{\mathtt{D},i}\lambda_{i}}{\mathcal{A}_{\mathtt{D},i}}y\exp\{-\pi\frac{q_{\mathtt{D},i}\lambda_{i}}{\mathcal{A}_{\mathtt{D},i}}y^{2}\},\label{eq:Dis_PDF_DL}
\end{equation}
\begin{equation}
f_{Y_{\mathtt{U},i}}(y)=2\pi\frac{\left(1-q_{\mathtt{D},i}\right)\lambda_{i}}{\mathcal{A}_{\mathtt{U},i}}y\exp\{-\pi\frac{\left(1-q_{\mathtt{D},i}\right)\lambda_{i}}{\mathcal{A}_{\mathtt{U},i}}y^{2}\},\label{eq:Dis_PDF_UL}
\end{equation}
where the result is a modification of Lemma 4 in \cite{HCNW} for
dynamic TDD networks.

The DL and UL SIR of a typical receiver associated with the macro
tier is given by
\begin{equation}
\mathtt{SIR}_{\mathtt{m}}^{\mathtt{D}}=\frac{P_{\mathtt{m}}h_{or}r^{-\alpha}}{\ensuremath{I_{\mathtt{D\rightarrow D}}^{\mathtt{(m)}}+I_{\mathtt{U\rightarrow D}}^{\mathtt{(m)}}}},\;\mathtt{SIR_{m}^{U}}=\frac{Q_{\mathtt{m}}h_{or}r^{-\alpha}}{\ensuremath{I_{\mathtt{D\rightarrow U}}^{\mathtt{(m)}}+I_{\mathtt{U\rightarrow U}}^{(\mathtt{m})}}},\label{eq:(15)}
\end{equation}
where $h_{or}$ and $r$ are the fading power and the typical link
length,%
\footnote{To clarify the channel of a specific link, $r$ in the subscript denotes
the position of a transmitter.%
} and
\[
I_{\mathtt{D\rightarrow D}}^{\mathtt{(m)}}=\sum_{y\in\Phi_{\mathtt{m}}^{\mathtt{D}}\backslash\{y_{0}\}}P_{\mathtt{m}}h_{oy}y^{-\alpha},\: I_{\mathtt{U\rightarrow D}}^{\mathtt{(m)}}=\sum_{x\in\Phi_{\mathtt{u,m}}^{\mathtt{T}}}Q_{\mathtt{m}}h_{ox}x^{-\alpha},
\]
\[
I_{\mathtt{D\rightarrow U}}^{\mathtt{(m)}}=\sum_{y\in\Phi_{\mathtt{m}}^{\mathtt{D}}}P_{\mathtt{m}}h_{oy}y^{-\alpha},\: I_{\mathtt{U\rightarrow U}}^{(\mathtt{m})}=\sum_{x\in\Phi_{\mathtt{u,m}}^{\mathtt{T}}\backslash\{x_{0}\}}Q_{\mathtt{m}}h_{ox}x^{-\alpha},
\]
where $y_{0}$ and $x_{0}$ represent the position of typical transmitter
in DL and UL mode, $\Phi_{\mathtt{u},\mathtt{m}}^{\mathtt{T}}$ represents
the set of transmitting mobile users associated with macro tier. Due
to the orthogonal multiple access technology, there is a one-to-one
mapping from the transmitting mobile users associated with macro tier
to the active UL MBSs. Since the coupling between the location of
MBSs and transmitting mobile users has little effect on the coverage
probability \cite{CHDD,AMUC}, we neglect the coupling and model $\Phi_{\mathtt{u},\mathtt{m}}^{\mathtt{T}}$
as a PPP with density $\lambda_{\mathtt{u,m}}^{\mathtt{T}}=\lambda_{\mathtt{m}}^{\mathtt{U}}$.
The simulation results in Section IV also validate the accuracy of
the approximation.

{\small{}}
\begin{figure}[t]
{\small{}\centering\includegraphics[scale=0.35]{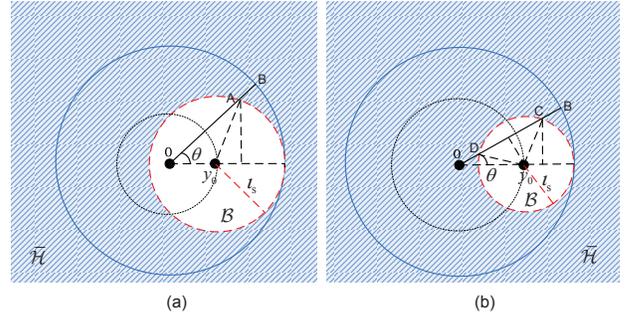}\protect\caption{{\small{}\label{fig:D2D_small cell}In the small cell tier, the typical
receiver locates at the origin, and the serving transmitter $y_{0}$
employs $\iota_{\mathtt{s}}$ to form an exclusion region, within
which no D2D transmitter can exist. (a) is for $\Vert y_{0}\Vert\leq\iota_{\mathtt{s}}$
and (b) is for $\Vert y_{0}\Vert>\iota_{\mathtt{s}}$. }}
}
\end{figure}
{\small \par}

The DL and UL SIR of a typical receiver associated with small cell
tier is denoted by
\[
\mathtt{SIR}_{\mathtt{s}}^{\mathtt{D}}=\frac{P_{\mathtt{s}}h_{or}r^{-\alpha}}{I_{\mathtt{D\rightarrow D}}^{\mathtt{(s)}}+I_{\mathtt{U\rightarrow D}}^{\mathtt{(s)}}+I_{\mathtt{d\rightarrow D}}},\:\mathtt{SIR}_{\mathtt{s}}^{\mathtt{U}}=\frac{Q_{\mathtt{s}}h_{or}r^{-\alpha}}{I_{\mathtt{D\rightarrow U}}^{\mathtt{(s)}}+I_{\mathtt{U\rightarrow U}}^{\mathtt{(s)}}+I_{\mathtt{d\rightarrow U}}},
\]
\begin{equation}
\label{eq:SIR_small}
\end{equation}
where
\[
I_{\mathtt{D\rightarrow D}}^{\mathtt{(s)}}=\sum_{y\in\Phi_{\mathtt{s}}^{\mathtt{D}}\backslash\{y_{0}\}}P_{\mathtt{s}}h_{oy}y^{-\alpha},\: I_{\mathtt{U\rightarrow D}}^{\mathtt{(s)}}=\sum_{x\in\Phi_{\mathtt{u},\mathtt{s}}^{\mathtt{T}}}Q_{\mathtt{s}}h_{ox}x^{-\alpha},
\]
\[
I_{\mathtt{d\rightarrow D}}=\sum_{z\in\Phi_{\mathtt{d}}\backslash b\left(y_{0},\iota_{\mathtt{s}}\right)}Q_{\mathtt{d}}h_{oz}z^{-\alpha},\: I_{\mathtt{D\rightarrow U}}^{\mathtt{(s)}}=\sum_{y\in\Phi_{\mathtt{s}}^{\mathtt{D}}}P_{\mathtt{s}}h_{oy}y^{-\alpha},
\]
\[
I_{\mathtt{U\rightarrow U}}^{\mathtt{(s)}}=\sum_{x\in\Phi_{\mathtt{u},\mathtt{s}}^{\mathtt{T}}\backslash\{x_{0}\}}Q_{\mathtt{s}}h_{ox}x^{-\alpha},\: I_{\mathtt{d\rightarrow U}}=\sum_{z\in\Phi_{\mathtt{d}}\backslash b\left(x_{0},\iota_{\mathtt{s}}\right)}Q_{\mathtt{d}}h_{oz}z^{-\alpha},
\]
where $\Phi_{\mathtt{u},\mathtt{s}}^{\mathtt{T}}\sim\textrm{PPP}(\lambda_{\mathtt{u,s}}^{\mathtt{T}})$
represents the set of transmitting mobile users associated with small
cell tier with density $\lambda_{\mathtt{u,s}}^{\mathtt{T}}=\lambda_{\mathtt{s}}^{\mathtt{U}}$.
As is shown in Fig. \ref{fig:D2D_small cell}, the active D2D transmitters
are distributed in the shaded region, i.e. the whole $\mathbb{R}^{2}$
plane except for the exclusion region centered at $y_{0}$, which
is approximated by a ball $\mathcal{B}$. The shaded region can be
divided into two disjoint parts by the circle $\mathcal{H}$ with
the center at the origin. Define $\mathcal{H}\overset{\triangle}{=}b\left(0,\Vert y_{0}\Vert+\iota_{\mathtt{s}}\right)$,
$\mathcal{B}\overset{\triangle}{=}b\left(y_{0},\iota_{\mathtt{s}}\right)$,
where $\mathcal{B}$ is the exclusion region for D2D transmissions
around each small cell transmitter. Note that as a function of $\rho_{\mathtt{s}}$
and the instantaneous channel fading, the exclusion region is not
a ball but an irregular shape, which varies in different timeslot.
To simplify the analysis, we use a ball to approximate the exclusion
region, where the equivalent exclusion distance is determined by imposing
a small miss detection probability threshold $\epsilon$. The constraint
is met when $\mathbb{\Pr}[Q_{\mathtt{d}}h_{x_{0}z}/\iota_{\mathtt{s}}^{\alpha}>\rho_{\mathtt{s}}]=\epsilon$,
and solving for $\iota_{\mathtt{s}}$ yields \textbf{$\iota_{\mathtt{s}}=\bigl(\frac{-\ln\epsilon}{\rho_{\mathtt{s}}/Q_{\mathtt{d}}}\bigr)^{\frac{1}{\alpha}}$},
where $h_{x_{0}z}$ denotes the fading power from a D2D transmitter
$z$ to the typical small cell transmitter $x_{0}$.

The SIR of a typical D2D receiver is given by
\begin{equation}
\mathtt{SIR_{d}}=\frac{Q_{\mathtt{d}}h_{or}r_{\mathtt{d}}^{-\alpha}}{I_{\mathtt{D\rightarrow d}}^{\mathtt{(s)}}+I_{\mathtt{U\rightarrow d}}^{\mathtt{(s)}}+I_{\mathtt{d\rightarrow d}}},
\end{equation}
where
\[
I_{\mathtt{D\rightarrow d}}^{\mathtt{(s)}}=\sum_{y\in\Phi_{\mathtt{s}}^{\mathtt{D}}\backslash b\left(z_{0},\iota_{\mathtt{s}}\right)}P_{\mathtt{s}}h_{oy}y^{-\alpha},\: I_{\mathtt{U\rightarrow d}}^{\mathtt{(s)}}=\sum_{x\in\Phi_{\mathtt{u},\mathtt{s}}^{\mathtt{T}}\backslash b\left(z_{0},\iota_{\mathtt{s}}\right)}Q_{\mathtt{s}}h_{ox}x^{-\alpha}
\]
\[
I_{\mathtt{d\rightarrow d}}=\sum_{z\in\Phi_{\mathtt{d}}\backslash b\left(z_{0},\iota_{\mathit{\mathtt{d}}}\right)}Q_{\mathtt{d}}h_{oz}z^{-\alpha}.
\]
There exists two exclusion regions $b\left(z_{0},\iota_{\mathtt{s}}\right)$
and $b\left(z_{0},\iota_{\mathtt{d}}\right)$ around each retained
D2D transmitter $z_{0}$, where the former one is due to the sensing
for small cell transmissions, and the latter one is resulted from
the sensing among D2D transmitters. Similar to the approximation for
the exclusion region around each small cell transmitter, the exclusion
regions around each retained D2D transmitter are also approximated
by two concentric balls with radius $\iota_{\mathtt{s}}$ and $\iota_{\mathtt{d}}$,
respectively. The radius $\iota_{\mathit{\mathtt{d}}}$ is constrained
by $\epsilon$ as $\Pr[Q_{\mathtt{d}}h_{z_{0}z}/\iota_{\mathtt{d}}^{\alpha}>\rho_{\mathtt{d}}]=\epsilon$,
and solving for $\iota_{\mathit{\mathtt{d}}}$ yields $\iota_{\mathtt{d}}=\bigl(\frac{-\ln\epsilon}{\rho_{\mathtt{d}}/Q_{\mathtt{d}}}\bigr)^{\frac{1}{\alpha}}.$

In the following lemma, we provide the retaining probability of each
potential D2D transmitter and derive the density of active D2D transmitters.

\begin{lemma}

The retaining probability of a potential D2D transmitter is given
by
\begin{equation}
\beta=\exp\left(-\left(\lambda_{\mathtt{s}}^{\mathtt{D}}+\lambda_{\mathtt{u,s}}^{\mathtt{T}}\right)\mathcal{K}_{o,\mathtt{s}}\right)\frac{1-\exp\left(-\zeta\lambda_{\mathtt{u}}\mathcal{K}_{o,\mathtt{d}}\right)}{\zeta\lambda_{\mathtt{u}}\mathcal{K}_{o,\mathtt{d}}},\label{eq:D2D_activity}
\end{equation}
and the corresponding density of active D2D transmitters is derived
as
\begin{equation}
\lambda_{\mathtt{d}}=\exp\left(-\left(\lambda_{\mathtt{s}}^{\mathtt{D}}+\lambda_{\mathtt{u,s}}^{\mathtt{T}}\right)\mathcal{K}_{o,\mathtt{s}}\right)\frac{1-\exp\left(-\zeta\lambda_{\mathtt{u}}\mathcal{K}_{o,\mathtt{d}}\right)}{\mathcal{K}_{o,\mathtt{d}}},\label{eq:D2D_dens}
\end{equation}
where $\mathcal{K}_{o,\mathtt{s}}=\frac{2\pi\Gamma\left(\frac{2}{\alpha}\right)}{\alpha\bigl(\frac{\rho_{\mathtt{s}}}{Q_{\mathtt{d}}}\bigr)^{\frac{2}{\alpha}}}$
and $\mathcal{K}_{o,\mathtt{d}}=\frac{2\pi\Gamma\left(\frac{2}{\alpha}\right)}{\alpha\bigl(\frac{\rho_{\mathtt{d}}}{Q_{\mathtt{d}}}\bigr)^{\frac{2}{\alpha}}}$.
\begin{IEEEproof}
See Appendix A.
\end{IEEEproof}
\end{lemma}

With the per tier association probability, we derive the overall coverage
probability of a mobile user associated with the infrastructure and
the coverage probability of a typical D2D receiver as follows.

\begin{theorem}

In a two-tier dynamic TDD heterogeneous network, the overall load-aware
coverage probability of a mobile user associated with the infrastructure
in DL and UL mode is given by
\begin{equation}
\bar{\mathbb{P}}_{\mathtt{D}}=\mathbb{P}_{\mathtt{m}}^{\mathtt{D}}\mathcal{A}_{\mathtt{D},\mathtt{m}}+\mathbb{P}_{\mathtt{s}}^{\mathtt{D}}\mathcal{A}_{\mathtt{D},\mathtt{s}},\:\bar{\mathbb{P}}_{\mathtt{U}}=\mathbb{P}_{\mathtt{m}}^{\mathtt{U}}\mathcal{A}_{\mathtt{U},\mathtt{m}}+\mathbb{P}_{\mathtt{s}}^{\mathtt{U}}\mathcal{A}_{\mathtt{U},\mathtt{s}},\label{eq:Ave_coverage}
\end{equation}
and the coverage probability of the typical D2D receiver is derived
as
\begin{equation}
\mathbb{P}_{\mathtt{d}}=\exp\biggl(-\mathcal{I}_{1}\Bigl(\frac{\gamma_{\mathtt{d}}r_{\mathtt{d}}^{\alpha}}{Q_{\mathtt{d}}};P_{\mathtt{s}}\Bigr)-\mathcal{I}_{2}\Bigl(\frac{\gamma_{\mathtt{d}}r_{\mathtt{d}}^{\alpha}}{Q_{\mathtt{d}}};Q_{\mathtt{s}}\Bigr)-\mathcal{I}_{3}\Bigl(\frac{\gamma_{\mathtt{d}}r_{\mathtt{d}}^{\alpha}}{Q_{\mathtt{d}}};Q_{\mathtt{d}}\Bigr)\biggr),\label{eq:D2D_coverage}
\end{equation}
where
\begin{eqnarray}
\mathbb{P}_{\mathtt{m}}^{\mathtt{D}} & = & \frac{q_{\mathtt{D,m}}\lambda_{\mathtt{m}}}{\lambda_{\mathtt{m}}^{\mathtt{D}}\mathcal{A}_{\mathtt{D}\mathtt{,m}}\delta\left(\gamma_{\mathtt{m}}^{\mathtt{D}},\alpha\right)+\lambda_{\mathtt{u,m}}^{\mathtt{T}}\mathcal{A}_{\mathtt{D}\mathtt{,m}}C\left(\alpha\right)\bigl(\frac{Q_{\mathtt{m}}}{P_{\mathtt{m}}}\gamma_{\mathtt{m}}^{\mathtt{D}}\bigr)^{\frac{2}{\alpha}}+q_{\mathtt{D,m}}\lambda_{\mathtt{m}}},\nonumber \\
\label{eq:macro_D}
\end{eqnarray}
\begin{eqnarray}
\mathbb{P}_{\mathtt{m}}^{\mathtt{U}} & = & \frac{\left(1-q_{\mathtt{D,m}}\right)\lambda_{\mathtt{m}}}{C\left(\alpha\right)\bigl(\gamma_{\mathtt{m}}^{\mathtt{U}}\bigr)^{\frac{2}{\alpha}}\mathcal{A}_{\mathtt{U}\mathtt{,m}}\Bigl(\lambda_{\mathtt{m}}^{\mathtt{D}}\bigl(\frac{P_{\mathtt{m}}}{Q_{\mathtt{m}}}\bigr)^{\frac{2}{\alpha}}+\lambda_{\mathtt{u,m}}^{\mathtt{T}}\Bigr)+\left(1-q_{\mathtt{D,m}}\right)\lambda_{\mathtt{m}}},\nonumber \\
\label{eq:macro_U}
\end{eqnarray}
\begin{eqnarray}
\mathbb{P}_{\mathtt{s}}^{\mathtt{D}} & = & \frac{\pi q_{\mathtt{D,s}}\lambda_{\mathtt{s}}}{\mathcal{A}_{\mathtt{D,s}}}\biggl[\int_{0}^{\iota_{\mathtt{s}}^{2}}e^{-\pi v\mathcal{F}}\mathcal{L}_{I_{\mathtt{d\rightarrow D}}}(\frac{\gamma_{\mathtt{s}}^{\mathtt{D}}v^{\frac{\alpha}{2}}}{P_{\mathtt{s}}}\mid v\leq\iota_{\mathtt{s}}^{2})dv\nonumber \\
 &  & +\int_{\iota_{\mathtt{s}}^{2}}^{\infty}e^{-\pi v\mathcal{F}}\mathcal{L}_{I_{\mathtt{d\rightarrow D}}}(\frac{\gamma_{\mathtt{s}}^{\mathtt{D}}v^{\frac{\alpha}{2}}}{P_{\mathtt{s}}}\mid v>\iota_{\mathtt{s}}^{2})dv\biggr],\label{eq:Small_D}
\end{eqnarray}
\begin{eqnarray}
\mathbb{P}_{\mathtt{s}}^{\mathtt{U}} & = & \frac{\pi\left(1-q_{\mathtt{D,s}}\right)\lambda_{\mathtt{s}}}{\mathcal{A}_{\mathtt{U,s}}}\biggl[\int_{0}^{\iota_{\mathtt{s}}^{2}}e^{-\pi v\mathcal{G}}\mathcal{L}_{I_{\mathtt{d\rightarrow U}}}(\frac{\gamma_{\mathtt{s}}^{\mathtt{U}}v^{\frac{\alpha}{2}}}{Q_{\mathtt{s}}}\mid v\leq\iota_{\mathtt{s}}^{2})dv\nonumber \\
 &  & +\int_{\iota_{\mathtt{s}}^{2}}^{\infty}e^{-\pi v\mathcal{G}}\mathcal{L}_{I_{\mathtt{d\rightarrow U}}}(\frac{\gamma_{\mathtt{s}}^{\mathtt{U}}v^{\frac{\alpha}{2}}}{Q_{\mathtt{s}}}\mid v>\iota_{\mathtt{s}}^{2})dv\biggr],\label{eq:Small_U}
\end{eqnarray}
\begin{eqnarray}
\mathcal{I}_{1}\left(s;P_{\mathtt{s}}\right) & = & \pi\lambda_{\mathtt{s}}^{\mathtt{D}}\bigl(\iota_{\mathtt{s}}+r_{\mathtt{d}}\bigr)^{2}\delta\Bigl(\frac{sP_{\mathtt{s}}}{\bigl(\iota_{\mathtt{s}}+r_{\mathtt{d}}\bigr)^{\alpha}},\alpha\Bigr)\nonumber \\
 &  & +\lambda_{\mathtt{s}}^{\mathtt{D}}\mathcal{Z}_{0,l_{OE}}^{\pi,\iota_{\mathtt{s}}+r_{\mathtt{d}}}\bigl(s;P_{\mathtt{s}}\bigr),\label{eq:I_1}
\end{eqnarray}
\begin{eqnarray}
\mathcal{I}_{2}\left(s;Q_{\mathtt{s}}\right) & = & \pi\lambda_{\mathtt{u,s}}^{\mathtt{T}}\bigl(\iota_{\mathtt{s}}+r_{\mathtt{d}}\bigr)^{2}\delta\Bigl(\frac{sQ_{\mathtt{s}}}{\bigl(\iota_{\mathtt{s}}+r_{\mathtt{d}}\bigr)^{\alpha}},\alpha\Bigr)\nonumber \\
 &  & +\lambda_{\mathtt{u,s}}^{\mathtt{T}}\mathcal{Z}_{0,l_{OE}}^{\pi,\iota_{\mathtt{s}}+r_{\mathtt{d}}}\bigl(s;Q_{\mathtt{s}}\bigr),\label{eq:I_2}
\end{eqnarray}
\begin{eqnarray}
\mathcal{I}_{3}\left(s;Q_{\mathtt{d}}\right) & = & \pi\lambda_{\mathtt{d}}\bigl(\iota_{\mathtt{d}}+r_{\mathtt{d}}\bigr)^{2}\delta\Bigl(\frac{sQ_{\mathtt{d}}}{\bigl(\iota_{\mathtt{d}}+r_{\mathtt{d}}\bigr)^{\alpha}},\alpha\Bigr)\nonumber \\
 &  & +\lambda_{\mathtt{d}}\mathcal{Z}_{0,l_{OF}}^{\pi,\iota_{\mathtt{d}}+r_{\mathtt{d}}}\bigl(s;Q_{\mathtt{d}}\bigr).\label{eq:I_3}
\end{eqnarray}
The variables in \prettyref{eq:D2D_coverage}-\prettyref{eq:I_3}
are defined as
\[
\mathcal{F\triangleq}\lambda_{\mathtt{s}}^{\mathtt{D}}\delta\bigl(\gamma_{\mathtt{s}}^{\mathtt{D}},\alpha\bigr)+\lambda_{\mathtt{u,s}}^{\mathtt{T}}C\left(\alpha\right)\bigl(\frac{Q_{\mathtt{s}}}{P_{\mathtt{s}}}\gamma_{\mathtt{s}}^{\mathtt{D}}\bigr)^{\frac{2}{\alpha}}+\frac{q_{\mathtt{D,s}}\lambda_{\mathtt{s}}}{\mathcal{A}_{\mathtt{D,s}}},
\]
\[
\mathcal{G\triangleq}C\left(\alpha\right)\bigl(\gamma_{\mathtt{s}}^{\mathtt{U}}\bigr)^{\frac{2}{\alpha}}\Bigl(\lambda_{\mathtt{s}}^{\mathtt{D}}\bigl(\frac{P_{\mathtt{s}}}{Q_{\mathtt{s}}}\bigr)^{\frac{2}{\alpha}}+\lambda_{\mathtt{u,s}}^{\mathtt{T}}\Bigr)+\frac{\left(1-q_{\mathtt{D,s}}\right)\lambda_{\mathtt{s}}}{\mathcal{A}_{\mathtt{U,s}}},
\]
\begin{eqnarray*}
\mathcal{L}_{I_{\mathtt{d\rightarrow D}}}(s\mid r\leq\iota_{\mathtt{s}}) & = & \mathcal{L}_{I_{\mathtt{d\rightarrow U}}}(s\mid r\leq\iota_{\mathtt{s}})\\
 & = & \mathcal{L}_{I_{out}}(s\mid r)\exp\Bigl(-\lambda_{\mathtt{d}}\mathcal{Z}_{0,l_{OA}}^{\pi,\iota_{s}+r}\left(s;Q_{\mathtt{d}}\right)\Bigr),
\end{eqnarray*}
\begin{eqnarray*}
\mathcal{L}_{I_{\mathtt{d\rightarrow D}}}(s\mid r>\iota_{\mathtt{s}}) & = & \mathcal{L}_{I_{\mathtt{d\rightarrow U}}}(s\mid r>\iota_{\mathtt{s}})\\
 & = & \mathcal{L}_{I_{out}}(s\mid r)\exp\Bigl(-\lambda_{\mathtt{d}}\bigl(\mathcal{Z}_{0,0}^{\Theta,l_{OD}}\left(s;Q_{\mathtt{d}}\right)\\
 &  & +\mathcal{Z}_{0,l_{OC}}^{\Theta,\iota_{\mathtt{s}}+r}\left(s;Q_{\mathtt{d}}\right)+\mathcal{Z}_{\Theta,0}^{\pi,\iota_{\mathtt{s}}+r}\left(s;Q_{\mathtt{d}}\right)\bigr)\Bigr),
\end{eqnarray*}
\[
\mathcal{L}_{I_{out}}(s\mid r)=\exp\Bigl(-\pi\lambda_{\mathtt{d}}\bigl(\iota_{\mathtt{s}}+r\bigr)^{2}\delta\Bigl(\frac{sQ_{\mathtt{d}}}{\bigl(\iota_{\mathtt{s}}+r\bigr)^{\alpha}},\alpha\Bigr)\Bigr),
\]
\[
\mathcal{Z}_{\theta_{l},\kappa_{l}}^{\theta_{u},\kappa_{u}}\left(s;Q\right)=\left(sQ\right)^{\frac{2}{\alpha}}\int_{\theta_{l}}^{\theta_{u}}\int_{\frac{\kappa_{l}^{2}}{(sQ)^{\frac{2}{\alpha}}}}^{\frac{\kappa_{u}^{2}}{\left(sQ\right)^{\frac{2}{\alpha}}}}\frac{1}{1+u^{\frac{\alpha}{2}}}dud\theta,
\]
{\small{}
\[
C\left(\alpha\right)=\frac{2\pi/\alpha}{\sin\left(2\pi/\alpha\right)},\;\delta\left(\beta,\alpha\right)=\int_{\beta^{-\frac{2}{\alpha}}}^{\infty}\frac{\beta^{\frac{2}{\alpha}}}{1+u^{\frac{\alpha}{2}}}du,
\]
\[
l_{OA}=\sqrt{\iota_{\mathtt{s}}^{2}-\bigl(r\textrm{sin}\theta\bigr)^{2}}+r\cos\theta,\; l_{OC}=\sqrt{\iota_{\mathtt{s}}^{2}-\left(r\textrm{sin}\theta\right)^{2}}-r\cos\theta,
\]
\[
\Theta=\arcsin\bigl(\frac{\iota_{\mathtt{s}}}{r}\bigr),\; l_{OD}=r\cos\theta-\sqrt{\iota_{\mathtt{s}}^{2}-\left(r\textrm{sin}\theta\right)^{2}},
\]
\[
l_{OE}=\sqrt{\iota_{\mathtt{s}}^{2}-\bigl(r_{\mathtt{d}}\textrm{sin}\theta\bigr)^{2}}+r_{\mathtt{d}}\cos\theta,\; l_{OF}=\sqrt{\iota_{\mathtt{d}}^{2}-\left(r_{\mathtt{d}}\textrm{sin}\theta\right)^{2}}+r_{\mathtt{d}}\cos\theta.
\]
}{\small \par}
\begin{IEEEproof}
The full proof is provided in Appendix B. In the above equations,
$\mathcal{F}$ and $\mathcal{G}$, respectively, correspond to the
interference inflicted by DL SAPs and transmitting mobile users on
the typical small cell receiver in DL and UL. $\mathcal{L}_{I_{\mathtt{d\rightarrow D}}}(s\mid r\leq\iota_{\mathtt{s}})$,
$\mathcal{L}_{I_{\mathtt{d\rightarrow D}}}(s\mid r>\iota_{\mathtt{s}})$
and $\mathcal{L}_{I_{\mathtt{d\rightarrow U}}}(s\mid r\leq\iota_{\mathtt{s}})$,
$\mathcal{L}_{I_{\mathtt{d\rightarrow U}}}(s\mid r>\iota_{\mathtt{s}})$
are the Laplace transforms of interference incurred by the active
D2D transmitters on the typical small cell receiver in DL and UL,
differentiated by the amplitude of the exclusion distance $\iota_{\mathtt{s}}$.
$\mathcal{I}_{1}(s;P_{\mathtt{s}})$, $\mathcal{I}_{2}(s;Q_{\mathtt{s}})$
and $\mathcal{I}_{3}(s;Q_{\mathtt{d}})$ correspond to the interference
incurred by the DL SAPs, transmitting mobile users and active D2D
transmitters, respectively.
\end{IEEEproof}
\end{theorem}

The adoption of the CSMA scheme in our analysis leads to elaborate
expressions of the coverage probability for the small cell tier. For
the most general case, it involves triple integrals that can be efficiently
solved by employing standard mathematical software packages. In the
following section, we present the asymptotic analysis related to the
protection threshold $\rho_{\mathtt{s}}$, which simplifies the analysis
substantially.

\subsection{Asymptotic Analysis}

\subsubsection{No D2D transmissions}

When $\rho_{\mathtt{s}}\rightarrow0$, we have $\lambda_{\mathtt{d}}\rightarrow0$.
The coverage probability of small cell tier in DL and UL is respectively,
\begin{equation}
\lim_{\rho_{\mathtt{s}}\rightarrow0}\mathbb{P}_{\mathtt{s}}^{\mathtt{D}}=\frac{q_{\mathtt{D,s}}\lambda_{\mathtt{s}}}{\lambda_{\mathtt{s}}^{\mathtt{D}}\mathcal{A}_{\mathtt{D}\mathtt{,s}}\delta\left(\gamma_{\mathtt{s}}^{\mathtt{D}},\alpha\right)+\lambda_{\mathtt{u,s}}^{\mathtt{T}}\mathcal{A}_{\mathtt{D}\mathtt{,s}}C\left(\alpha\right)\bigl(\frac{Q_{\mathtt{s}}}{P_{\mathtt{s}}}\gamma_{\mathtt{s}}^{\mathtt{D}}\bigr)^{\frac{2}{\alpha}}+q_{\mathtt{D,s}}\lambda_{\mathtt{s}}}\label{eq:Cov_s_0_D}
\end{equation}
\begin{equation}
\lim_{\rho_{\mathtt{s}}\rightarrow0}\mathbb{P}_{\mathtt{s}}^{\mathtt{U}}=\frac{\left(1-q_{\mathtt{D,s}}\right)\lambda_{\mathtt{s}}}{C\left(\alpha\right)\left(\gamma_{\mathtt{s}}^{\mathtt{U}}\right)^{\frac{2}{\alpha}}\mathcal{A}_{\mathtt{U}\mathtt{,s}}\bigl(\lambda_{\mathtt{s}}^{\mathtt{D}}\bigl(\frac{P_{\mathtt{s}}}{Q_{\mathtt{s}}}\bigr)^{\frac{2}{\alpha}}+\lambda_{\mathtt{u,s}}^{\mathtt{T}}\bigr)+\left(1-q_{\mathtt{D,s}}\right)\lambda_{\mathtt{s}}}\label{eq:Cov_s_0_U}
\end{equation}
\textit{\emph{In the absence of D2D transmissions}}, the overall coverage
probability can be found by inserting \prettyref{eq:macro_D}, \prettyref{eq:macro_U},
\prettyref{eq:Cov_s_0_D} and \prettyref{eq:Cov_s_0_U} into \prettyref{eq:Ave_coverage}.

\subsubsection{No sensing for small cell transmissions }

When $\rho_{\mathtt{s}}\rightarrow\infty$, the active D2D transmitters
form an MHP with the retaining probability $\beta=\frac{1-\exp\left(-\zeta\lambda_{\mathtt{u}}\mathcal{K}_{o,\mathtt{d}}\right)}{\zeta\lambda_{\mathtt{u}}\mathcal{K}_{o,\mathtt{d}}}\overset{(a)}{\approx}\frac{1}{\zeta\lambda_{\mathtt{u}}\mathcal{K}_{o,\mathtt{d}}}$,
where (a) comes from $\zeta\lambda_{\mathtt{u}}\mathcal{K}_{o,\mathtt{d}}\gg1.$
Thereby the density of active D2D transmitters is given by $\lambda_{\mathtt{d}}=\beta\zeta\lambda_{\mathtt{u}}\approx\frac{1}{\mathcal{K}_{o,\mathtt{d}}}=\frac{\alpha\bigl(\frac{\rho_{\mathtt{d}}}{Q_{\mathtt{d}}}\bigr)^{\frac{2}{\alpha}}}{2\pi\Gamma\left(\frac{2}{\alpha}\right)}$.
The coverage probability of small cell tier in DL and UL is respectively
given by \prettyref{eq:small_limit} and \prettyref{eq:small_limit_U}
at the top of the next page. Accordingly, \textit{\emph{without sensing
for small cell transmissions}}, we can derive the overall coverage
probability by inserting \prettyref{eq:macro_D}, \prettyref{eq:macro_U},
\prettyref{eq:small_limit} and \prettyref{eq:small_limit_U} into
\prettyref{eq:Ave_coverage}.

\begin{figure*}[tbh]
\begin{eqnarray}
\lim_{\rho_{\mathtt{s}}\rightarrow\infty}\mathbb{P}_{\mathtt{s}}^{\mathtt{D}} & \approx & \frac{q_{\mathtt{D,s}}\lambda_{\mathtt{s}}}{\lambda_{\mathtt{s}}^{\mathtt{D}}\mathcal{A}_{\mathtt{D}\mathtt{,s}}\delta\left(\gamma_{\mathtt{s}}^{\mathtt{D}},\alpha\right)+C\left(\alpha\right)\bigl(\gamma_{\mathtt{s}}^{\mathtt{D}}\bigr)^{\frac{2}{\alpha}}\mathcal{A}_{\mathtt{D}\mathtt{,s}}\Bigl(\lambda_{\mathtt{u,s}}^{\mathtt{T}}\bigl(\frac{Q_{\mathtt{s}}}{P_{\mathtt{s}}}\bigr)^{\frac{2}{\alpha}}+\frac{\alpha\bigl(\frac{\rho_{\mathtt{d}}}{P_{\mathtt{s}}}\bigr)^{\frac{2}{\alpha}}}{2\pi\Gamma\left(\frac{2}{\alpha}\right)}\Bigr)+q_{\mathtt{D,s}}\lambda_{\mathtt{s}}},\label{eq:small_limit}
\end{eqnarray}
\begin{eqnarray}
\lim_{\rho_{\mathtt{s}}\rightarrow\infty}\mathbb{P}_{\mathtt{s}}^{\mathtt{U}} & \approx & \frac{\left(1-q_{\mathtt{D,s}}\right)\lambda_{\mathtt{s}}}{C\left(\alpha\right)\left(\gamma_{\mathtt{s}}^{\mathtt{U}}\right)^{\frac{2}{\alpha}}\mathcal{A}_{\mathtt{U}\mathtt{,s}}\Bigl(\lambda_{\mathtt{s}}^{\mathtt{D}}\bigl(\frac{P_{\mathtt{s}}}{Q_{\mathtt{s}}}\bigr)^{\frac{2}{\alpha}}+\lambda_{\mathtt{u,s}}^{\mathtt{T}}+\frac{\alpha\bigl(\frac{\rho_{\mathtt{d}}}{Q_{\mathtt{s}}}\bigr)^{\frac{2}{\alpha}}}{2\pi\Gamma\left(\frac{2}{\alpha}\right)}\Bigr)+\left(1-q_{\mathtt{D,s}}\right)\lambda_{\mathtt{s}}}.\label{eq:small_limit_U}
\end{eqnarray}
\end{figure*}

\subsection{Network throughput }

With the coverage probability obtained in Theorem 1, we derive the
sum throughput of the two-tier network, where the bandwidth of each
tier is normalized by $W$. We consider outage capacity with constant
bit-rate coding, such that the total network throughput in DL and
UL mode can be written as
\begin{equation}
\mathcal{T}_{\mathtt{D}}\left(\eta;\rho_{\mathtt{s}};\rho_{\mathtt{d}}\right)=\eta\mathcal{T}_{\mathtt{m}}^{\mathtt{D}}+\left(1-\eta\right)(\mathcal{T}_{\mathtt{s}}^{\mathtt{D}}+\frac{1}{2}\mathcal{T}_{\mathtt{d}}\bigr),\label{eq:thrpt_D}
\end{equation}
\begin{equation}
\mathcal{T}_{\mathtt{U}}\left(\eta;\rho_{\mathtt{s}};\rho_{\mathtt{d}}\right)=\eta\mathcal{T}_{\mathtt{m}}^{\mathtt{U}}+\left(1-\eta\right)(\mathcal{T}_{\mathtt{s}}^{\mathtt{U}}+\frac{1}{2}\mathcal{T}_{\mathtt{d}}\bigr).\label{eq:thrpt_U}
\end{equation}
where $\mathcal{T}_{i}^{\mathtt{D}}=\lambda_{i}^{\mathtt{D}}\mathbb{P}_{i}^{\mathtt{D}}\log_{2}(1+\gamma_{i}^{\mathtt{D}})$,
$\mathcal{T}_{i}^{\mathtt{U}}=\lambda_{\mathtt{u},i}^{\mathtt{T}}\mathbb{P}_{i}^{\mathtt{U}}\log_{2}(1+\gamma_{i}^{\mathtt{U}})$
and $\mathcal{T}_{\mathtt{d}}=\lambda_{\mathtt{d}}\mathbb{P}_{\mathtt{d}}\log_{2}(1+\gamma_{\mathtt{d}})$.
Half of the D2D outage capacity is included in the DL and UL network
throughput, respectively. Note that in \prettyref{eq:thrpt_D} and
\prettyref{eq:thrpt_U}, the load of base stations is incorporated
in the calculation of \emph{active} transmitter density $\lambda_{i}^{\mathtt{D}}$
and $\lambda_{\mathtt{u},i}^{\mathtt{T}}$ with the empty cells being
excluded.

\subsection{Validation}

In this section, we verify by means of simulations the validity of
the theoretical model and the approximations therein made concerning
the active D2D transmitters and the active transmitting mobile users.
All simulations are performed over a square window of 5000 $\times$
5000 $\textrm{m}^{\textrm{2}}$ with 10000 iterations. Unless otherwise
specified, we use the default values of the system parameters as shown
in Table I at the top of the next.
\begin{table*}[t]
{\Large{}\protect\caption{\label{tab:Notation-and-default}Notation and default values.}
}{\Large \par}

\centering%
\begin{tabular}{|c|c|c|}
\hline
{\small{}Notation} & {\small{}Description} & {\small{}Default Value}\tabularnewline
\hline
\hline
$\alpha$ & {\small{}Path loss exponent} & {\small{}4}\tabularnewline
\hline
$\lambda_{\mathtt{m}}$ & {\small{}Density of MBSs $\Phi_{\mathtt{m}}$} & {\small{}$1/(\pi500^{2})$ $\textrm{\ensuremath{m^{-2}}}$}\tabularnewline
\hline
$\lambda_{\mathtt{s}}$ & {\small{}Density of SAPs $\Phi_{\mathtt{s}}$ } & {\small{}Scenario dependent}\tabularnewline
\hline
$\lambda_{\mathtt{u}}$ & {\small{}Density of mobile users $\Phi_{\mathtt{u}}$} & {\small{}Scenario dependent}\tabularnewline
\hline
$P_{\mathtt{m}}$, $Q_{\mathtt{m}}$ & {\small{} DL, UL Transmit power, Macro tier} & {\small{}46 dBm, 20dBm}\tabularnewline
\hline
$P_{\mathtt{s}}$, $Q_{\mathtt{s}}$ & {\small{}DL, UL Transmit Power, Small cell tier} & {\small{}26 dBm, 10dBm}\tabularnewline
\hline
$Q_{\mathtt{d}}$  & {\small{}Transmit Power of D2D user} & {\small{}0 dBm}\tabularnewline
\hline
$\gamma_{\mathtt{m}}^{\mathtt{D}}$, $\gamma_{\mathtt{m}}^{\mathtt{U}}$ & {\small{}DL, UL SIR Threshold, Macro tier} & {\small{}0 dB, 0 dB}\tabularnewline
\hline
$\gamma_{\mathtt{s}}^{\mathtt{D}}$, $\gamma_{\mathtt{s}}^{\mathtt{U}}$ & {\small{}DL, UL SIR Threshold, Small cell tier} & {\small{}0 dB, 0 dB}\tabularnewline
\hline
$\gamma_{\mathtt{d}}$ & {\small{}D2D user Threshold} & {\small{}0 dB}\tabularnewline
\hline
$r_{\mathtt{d}}$ & {\small{}D2D link length} & {\small{}20 $\textrm{m}$}\tabularnewline
\hline
$\rho_{\mathtt{s}}$ & {\small{}Protection Threshold } & {\small{}-60 dBm}\tabularnewline
\hline
$\rho_{\mathtt{d}}$ & {\small{}Contention Threshold} & {\small{}-60 dBm}\tabularnewline
\hline
$\eta$ & {\small{}Bandwidth partition } & {\small{}Scenario dependent}\tabularnewline
\hline
$\zeta$  & {\small{}D2D transmitter fraction } & {\small{}Scenario dependent}\tabularnewline
\hline
$\mu$ & {\small{}Transmitting Mobile users fraction } & {\small{}0.5}\tabularnewline
\hline
$\epsilon$ & {\small{}Threshold of miss detection probability} & {\small{}$10^{-5}$}\tabularnewline
\hline
\end{tabular}
\end{table*}
\begin{figure}[t]
\centering\includegraphics[scale=0.35]{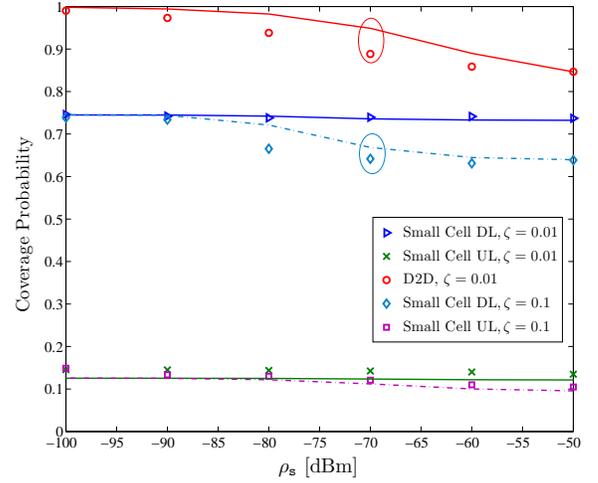}

\protect\caption{\label{fig:D2D_Small_cove_rho_s_d}Comparison of D2D and small cell
tier coverage probability from simulation (markers) and theoretical
analysis (lines) as a function of $\rho_{\mathtt{s}}$, for $\lambda_{\mathtt{s}}=5\lambda_{\mathtt{m}}$,
$\lambda_{\mathtt{u}}=100\lambda_{\mathtt{m}}$, $\{q_{\mathtt{D,m}},q_{\mathtt{D,s}}\}=\{0.5,0.5\}$,
$\rho_{\mathtt{d}}=-60$ dBm, and $\{B_{\mathtt{D,m}},B_{\mathtt{D,s}},B_{\mathtt{U,m}},B_{\mathtt{U,s}}\}=\{1,1,1,1\}$.}
\end{figure}

In Fig. \ref{fig:D2D_Small_cove_rho_s_d}, we study the validity of
the PPP approximation of active D2D transmitters in terms of coverage
probability by varying $\rho_{\mathtt{s}}$. The approximation is
caused by the following factors: (i) modeling the combined effect
of PHP and MHP with an independent thinning of a PPP, (ii) neglecting
the coupling between the locations of mobile users and base stations
in the UL transmission, (iii) replacing the instantaneous exclusion
distance by $\iota_{\mathtt{s}}$ and $\iota_{\mathtt{d}}$ constrained
by a small miss detection probability threshold $\epsilon$. Figure
\ref{fig:D2D_Small_cove_rho_s_d} indicates that the PPP approximation
is accurate for low and high values of $\rho_{\mathtt{s}}$. Low values
of $\rho_{\mathtt{s}}$ correspond to large exclusion distances $\iota_{\mathtt{s}}$,
leading to a low retaining probability $\beta$. The good agreement
between simulation and analysis can be explained by the fact that
the smaller density of D2D transmitters leads to little interference.
For high values of $\rho_{\mathtt{s}}$ and corresponding small exclusion
distances $\iota_{\mathtt{s}}$, the density of the active D2D transmitters
approaches that of the initial PPP, which eliminates the inaccuracy
caused by the approximation. The middle range of values of $\rho_{\mathtt{s}}$
results in inaccuracy on the coverage probability. Specifically, define
the receiver sensitivity as $\rho_{\mathtt{min}}$, the extensive
simulations show that the approximations are accurate (with the order
of magnitude of the error less than 5\%), when $\rho_{\mathtt{s}}\in[\rho_{\mathtt{min}},-95]\cup[-60,\infty)$
dBm, and reasonable (error order is within 10\%) within the range
$(-95,-60)$ dBm. In this example, the maximum error is achieved at
$\rho_{\mathtt{s}}=-70$ dBm. However, the order of magnitude of the
largest error is less than 10\%. Similar effect can be seen for $\rho_{\mathtt{d}}$,
and we find that the PPP approximation by varying $\rho_{\mathtt{d}}$
is more accurate than by varying $\rho_{\mathtt{s}}$. This can be
explained by the larger effect of $\rho_{\mathtt{s}}$ on $\lambda_{\mathtt{d}}$
than the effect of $\rho_{\mathtt{d}}$ on $\lambda_{\mathtt{d}}$.

\begin{figure}[t]
\centering\includegraphics[scale=0.3]{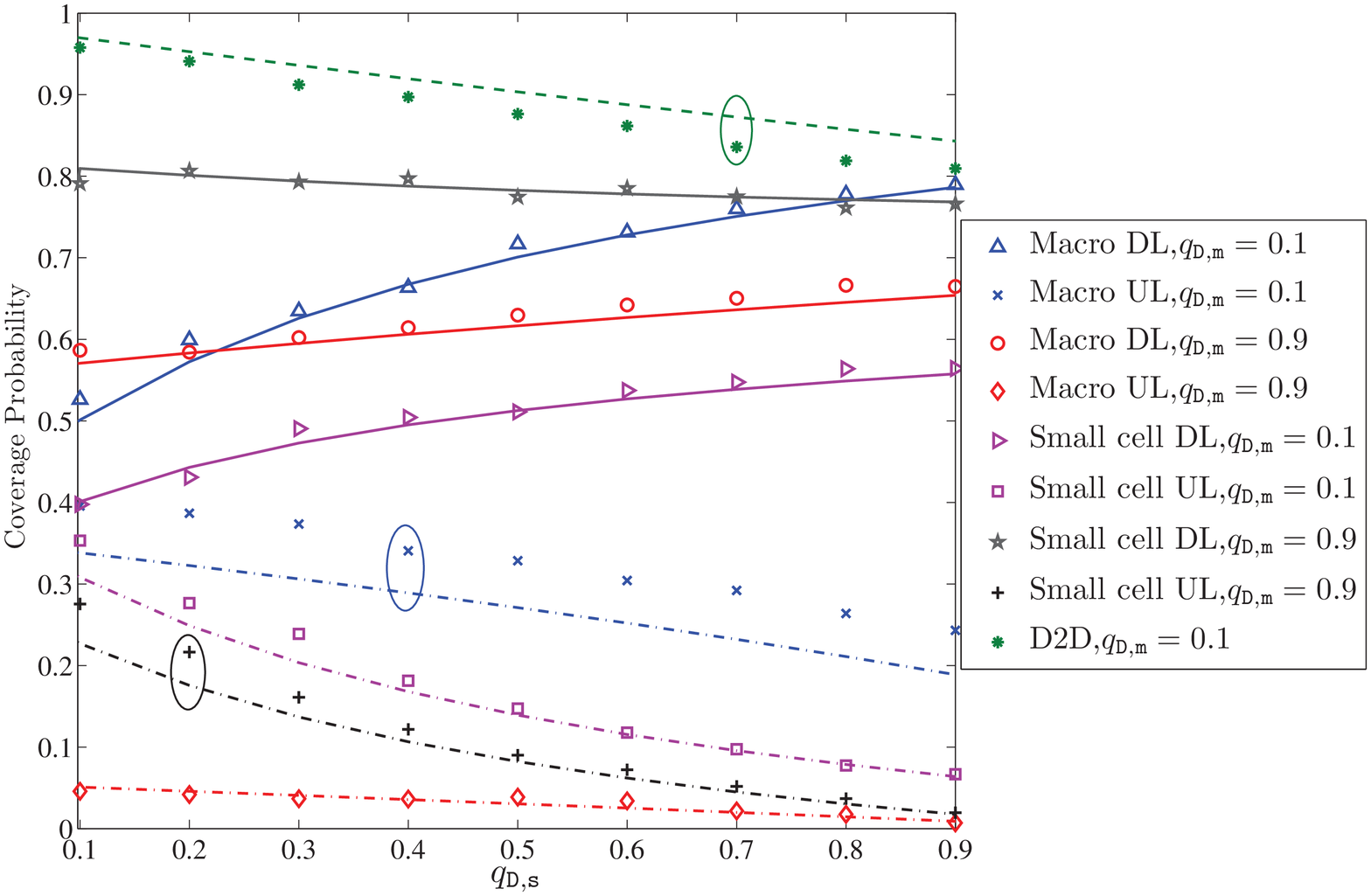}

\protect\caption{\label{fig:Cove_q_ds}Comparison of coverage probability from simulation
(markers) and theoretical analysis (lines) as a function of $q_{\mathtt{D},\mathtt{s}}$,
($\lambda_{\mathtt{s}}=5\lambda_{\mathtt{m}}$, $\lambda_{\mathtt{u}}=100\lambda_{\mathtt{m}}$,
$\{B_{\mathtt{D,m}},B_{\mathtt{D,s}},B_{\mathtt{U,m}},B_{\mathtt{U,s}}\}=\{1,1,1,1\}$,
$\zeta=0.1$, $\rho_{\mathtt{s}}=\rho_{\mathtt{d}}=-60$ dBm).}
\end{figure}

Figure \ref{fig:Cove_q_ds} represents the coverage probability as
a function of $q_{\mathtt{D,s}}$. We observe that the accuracy of
the approximation deteriorates as more UL transmissions take place.
For D2D users, the inaccuracy is mainly caused by the use of fixed
$\iota_{\mathtt{s}}$, where the channel fading is averaged. Note
that the D2D user coverage is dominated by the strong interference
from DL active SAPs. As $q_{\mathtt{D,s}}$ increases, the density
of DL SAPs increases and the impact of channel fading on the approximation
is more apparent. To further verify the approximations, we perform
extensive simulations by varying the related system parameters. Specifically,
by increasing $\zeta$ or $\lambda_{\mathtt{s}}$, we observe a larger
error of the approximation with regards to the D2D user coverage.
In the worst case with $\zeta=0.9$, even in a sparse network scenario
$\hat{\lambda}_{\mathtt{s}}^{(\mathtt{m})}=5$, the order of magnitude
of the error can achieve 10\%. While with a smaller D2D user fraction
$\zeta=0.1$, the approximation can be accurate (error order is within
5\%) in both sparse and moderate network scenario with $\hat{\lambda}_{\mathtt{s}}^{(\mathtt{m})}\leq40$.
However, in a very dense network scenario with $\hat{\lambda}_{\mathtt{s}}^{(\mathtt{m})}=100$,
the order of magnitude of the error can be as large as 16\%. The reasonable
order of magnitude of the worst-case errors validates our theoretical
model and in the following, we will present results based on our analytical
framework.

\section{Coverage Probability Evaluation}

In this section, we evaluate how the important network parameters,
such as UL/DL configuration, base station density, and bias factor
affect the load-aware coverage probability. In case of a fully-loaded
network without D2D users, we derive the parameters that maximize
the per tier coverage probability and overall coverage probability.

\subsection{Effects of UL/DL configuration}

In this section, we elucidate the non-trivial system behavior in dynamic
TDD networks that results from the coexistence of UL and DL transmissions.
How the UL/DL configuration $q_{\mathtt{D},i}$ affects\emph{ }$\mathbb{P}_{i}^{\mathtt{D}}$
and $\mathbb{P}_{i}^{\mathtt{U}}$ is not very explicit, because increasing
$q_{\mathtt{D},i}$ gives rise to a reduction of the UL interference
and a surge of the DL interference. However, for a given set of system
parameters, we find for each tier $i$ that the relative transmit
power $\frac{Q_{i}}{P_{i}}$ determines whether $\mathbb{P}_{i}^{\mathtt{D}}$
is dominated by the DL or the UL interference. According to this observation,
in a fully-loaded network with $\rho_{\mathtt{s}}\rightarrow0$, we
derive the UL/DL configuration that optimizes the per tier coverage
probability in DL and UL mode.

\emph{Optimization of per tier UL/DL configuration:} In a fully-loaded
network, i.e. $P_{e}^{\mathtt{D},k}\rightarrow0$ and $P_{e}^{\mathtt{U},k}\rightarrow0$,
and for $\rho_{\mathtt{s}}\rightarrow0$, the UL/DL configuration
that maximizes the per tier UL coverage probability $\mathbb{P}_{\mathtt{m}}^{\mathtt{U}}$
and $\mathbb{P}_{\mathtt{s}}^{\mathtt{U}}$ is given by $q_{\mathtt{D,m}}^{\star\mathtt{U}}=0$
and $q_{\mathtt{D,s}}^{\star\mathtt{U}}=0$, respectively. The optimal
UL/DL configuration for the per tier DL coverage probability is derived
as follows. Define $\bar{q}_{\mathtt{D},k}=\hat{\lambda}_{i}^{(k)}\bigl(\hat{P}_{i}^{(k)}\hat{B}_{\mathtt{D},i}^{(k)}\bigr)^{\frac{2}{\alpha}}\Bigl(\frac{\delta\left(\gamma_{i}^{\mathtt{D}},\alpha\right)}{C(\alpha)\bigl(\frac{Q_{i}}{P_{i}}\gamma_{i}^{\mathtt{D}}\bigr)^{\frac{2}{\alpha}}}-1\Bigr)^{-1},\: k,i\in\{\mathtt{m},\mathtt{s}\},k\neq i.$

(i) If $\frac{\delta\left(\gamma_{i}^{\mathtt{D}},\alpha\right)}{C(\alpha)\bigl(\gamma_{i}^{\mathtt{D}}\bigr)^{\frac{2}{\alpha}}}\leq\bigl(\frac{Q_{i}}{P_{i}}\bigr)^{\frac{2}{\alpha}}$,
$\mathbb{P}_{i}^{\mathtt{D}}$ is a monotone increasing function of
$q_{\mathtt{D},i}$, where $\mathbb{P}_{i}^{\mathtt{D}}$ is dominated
by the UL interference. The optimal UL/DL configuration is achieved
at $q_{\mathtt{D,}i}^{\star\mathtt{D}}=1$;

(ii) If $\frac{\delta\left(\gamma_{i}^{\mathtt{D}},\alpha\right)}{C(\alpha)\bigl(\gamma_{i}^{\mathtt{D}}\bigr)^{\frac{2}{\alpha}}}>\bigl(\frac{Q_{i}}{P_{i}}\bigr)^{\frac{2}{\alpha}}$,
the monotonicity of $\mathbb{P}_{i}^{\mathtt{D}}$ with respect to
$q_{\mathtt{D},i}$ is determined by the range of $q_{\mathtt{D},k}$.
When $q_{\mathtt{D},k}<\bar{q}_{\mathtt{D},k}$, $\mathbb{P}_{i}^{\mathtt{D}}$
increases with $q_{\mathtt{D},i}$, and we have $q_{\mathtt{D,}i}^{\star\mathtt{D}}=1$;
when $q_{\mathtt{D},k}>\bar{q}_{\mathtt{D},k}$, $\mathbb{P}_{i}^{\mathtt{D}}$
is a decreasing function of $q_{\mathtt{D},i}$, where $\mathbb{P}_{i}^{\mathtt{D}}$
is dominated by the DL interference. The optimal UL/DL configuration
is achieved at the limiting case of $q_{\mathtt{D,}i}^{\star\mathtt{D}}=0$.

The results can be obtained by taking the the first-order derivative
of $\mathbb{P}_{i}^{\mathtt{D}}$ and $\mathbb{P}_{i}^{\mathtt{U}}$
with respect to $q_{\mathtt{D},i}$. In a realistic scenario, the
base station has larger transmit power than that of mobile user, i.e.,
$P_{i}>Q_{i}$. Therefore, we have $\frac{\partial\mathbb{P}_{i}^{\mathtt{U}}}{\partial q_{\mathtt{D},i}}<0$,
which means that $\mathbb{P}_{i}^{\mathtt{U}}$ decreases with $q_{\mathtt{D},i}$.

\subsection{Effects of base station density}

\begin{figure}[t]
\subfloat[]{\centering\includegraphics[scale=0.35]{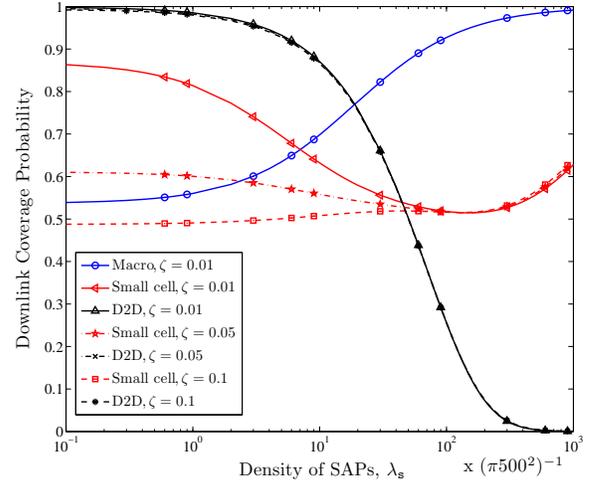}}

\subfloat[]{\centering\includegraphics[scale=0.35]{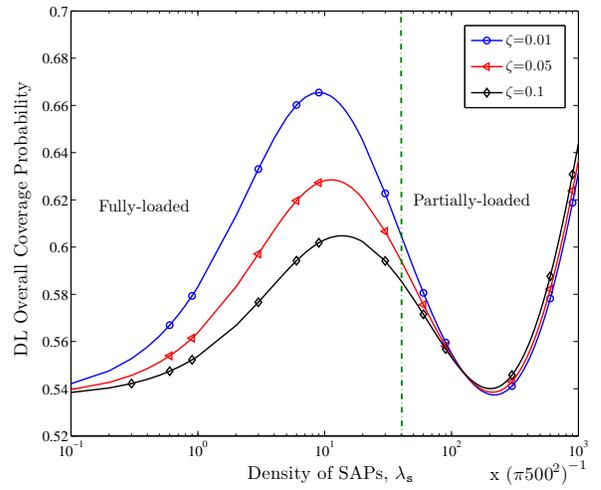}}

\protect\caption{\label{fig:Ave_Cove_dense}Downlink coverage probability vs. $\lambda_{\mathtt{s}}$,
(a) for macro tier, small cell tier and D2D user, (b) for overall
coverage of the infrastructure, ($\lambda_{\mathtt{m}}=\frac{1}{\pi500^{2}}$,
$\lambda_{\mathtt{u}}=10^{3}\lambda_{\mathtt{m}}$, $\{q_{\mathtt{D,m}},q_{\mathtt{D,s}}\}=\{0.5,0.5\}$,
$\{B_{\mathtt{D,m}},B_{\mathtt{D,s}},B_{\mathtt{U,m}},B_{\mathtt{U,s}}\}=\{1,1,1,1\}$,
$\rho_{\mathtt{s}}=\rho_{\mathtt{d}}=-60$ dBm). }
\end{figure}

From \prettyref{eq:macro_D} to \prettyref{eq:Small_U} and the definition
of $\mathcal{A}_{\mathtt{D},k}$ and $\mathcal{A}_{\mathtt{U},k}$,
we observe that both the base station density and bias factor have
similar effects on the coverage probability in DL and UL. Due to space
limitations, we take DL coverage as an example and the conclusions
can be directly applied to the UL case. We evaluate the variation
of the DL load-aware coverage probability as a function of $\lambda_{\mathtt{s}}$,
as depicted in Fig. \ref{fig:Ave_Cove_dense}. In terms of traffic
load, the network evolves from a fully-loaded sparse network to a
partially-loaded dense network. From Fig. \ref{fig:Ave_Cove_dense}(a),
we observe that $\mathbb{P}_{\mathtt{m}}^{\mathtt{D}}$ increases
monotonously with $\lambda_{\mathtt{s}}$, which can be ascribed to
the handover of macro mobile users with low SIR to the small cell
tier and the corresponding reduction of interference in the macro
tier. With respect to the small cell tier, the small cell network
interference increases with $\lambda_{\mathtt{s}}$, while the activity
of D2D users diminishes exponentially with $\lambda_{\mathtt{s}}$
as can be verified in \prettyref{eq:D2D_activity}. These opposite
effects are reflected in the load-aware coverage probability for the
small cell tier. Figure \ref{fig:Ave_Cove_dense}(b) depicts the overall
DL coverage probability $\bar{\mathbb{P}}_{\mathtt{D}}$ as a function
of $\lambda_{\mathtt{s}}$ and indicates that an optimal $\lambda_{\mathtt{s}}$
can be found in the feasible region of small cell densities. As $\lambda_{\mathtt{s}}$
increases, the network load moves into the lightly loaded regime,
where the aggregate small cell interference is constrained by the
density of mobile users. As $\lambda_{\mathtt{s}}\rightarrow\infty$,
we have $\mathcal{A}_{\mathtt{D},\mathtt{s}}\rightarrow1$ and $\bar{\mathbb{P}}_{\mathtt{D}}\rightarrow1$.
As opposed to the fully-loaded traffic model with constant coverage
probability in the asymptotic regime \cite{ATAT,MAKT}, this result
highlights the usefulness of the load-aware model to capture the coverage
probability in realistic conditions. Given a good estimate of the
user density, the proposed analytical framework allows us to find
the small cell density within the realistic regime that optimizes
the overall coverage probability. In the fully-loaded network and
for $\rho_{\mathtt{s}}\rightarrow0$, by taking the first-order derivative
of $\bar{\mathbb{P}}_{\mathtt{D}}$ and $\bar{\mathbb{P}}_{\mathtt{U}}$
with respect to $\hat{\lambda}_{\mathtt{s}}^{(\mathtt{m})}$, the
optimal relative base station density in DL mode $\hat{\lambda}_{\mathtt{s}}^{\mathtt{\left(m\right)\star D}}$
and UL mode $\hat{\lambda}_{\mathtt{s}}^{\mathtt{\left(m\right)\star U}}$
can be found as follows.

\emph{Optimization of base station density}: In the fully-loaded network,
the optimal $\hat{\lambda}_{\mathtt{s}}^{\mathtt{\left(m\right)\star D}}$
and $\hat{\lambda}_{\mathtt{s}}^{\mathtt{\left(m\right)\star U}}$
are given by
\begin{equation}
\hat{\lambda}_{\mathtt{s}}^{(\mathtt{m})\star\mathtt{D}}=\frac{q_{\mathtt{D,m}}P_{\mathtt{m}}^{\frac{2}{\alpha}}\delta(\gamma_{\mathtt{m}}^{\mathtt{D}},\alpha)+(1-q_{\mathtt{D},\mathtt{m}})\bigl(Q_{\mathtt{m}}\gamma_{\mathtt{m}}^{\mathtt{D}}\bigr)^{\frac{2}{\alpha}}}{\bigl(\hat{B}_{\mathtt{D,s}}^{(\mathtt{m})}\bigr)^{\frac{2}{\alpha}}\bigl(q_{\mathtt{D,s}}P_{\mathtt{s}}^{\frac{2}{\alpha}}\delta(\gamma_{\mathtt{s}}^{\mathtt{D}},\alpha)+(1-q_{\mathtt{D},\mathtt{s}})C(\alpha)\bigl(Q_{\mathtt{s}}\gamma_{\mathtt{s}}^{\mathtt{D}}\bigr)^{\frac{2}{\alpha}}\bigr)}\label{eq:Op_dens_D}
\end{equation}
\begin{equation}
\hat{\lambda}_{\mathtt{s}}^{(\mathtt{m})\star\mathtt{U}}=\Bigl(\frac{\gamma_{\mathtt{m}}^{\mathtt{U}}}{\hat{B}_{\mathtt{U,s}}^{(\mathtt{m})}\gamma_{\mathtt{s}}^{\mathtt{U}}}\Bigr)^{\frac{2}{\alpha}}\cdot\frac{q_{\mathtt{D},\mathtt{m}}P_{\mathtt{m}}^{\frac{2}{\alpha}}+(1-q_{\mathtt{D},\mathtt{m}})Q_{\mathtt{m}}^{\frac{2}{\alpha}}}{q_{\mathtt{D},\mathtt{s}}P_{\mathtt{s}}^{\frac{2}{\alpha}}+(1-q_{\mathtt{D},\mathtt{s}})Q_{\mathtt{s}}^{\frac{2}{\alpha}}},\label{eq:Op_dens_U}
\end{equation}

By analyzing the effect of key parameters, we can derive the following
insights: (i) the optimal base station density $\hat{\lambda}_{\mathtt{s}}^{(\mathtt{m})\star\mathtt{D}}$
and $\hat{\lambda}_{\mathtt{s}}^{(\mathtt{m})\star\mathtt{U}}$ decreases
with $\hat{B}_{\mathtt{D,s}}^{(\mathtt{m})}$ and $\hat{B}_{\mathtt{U,s}}^{(\mathtt{m})}$,
respectively, (ii) for DL case, if $P_{\mathtt{m}}\gg Q_{\mathtt{m}}$
and $P_{\mathtt{s}}\gg Q_{\mathtt{s}}$, $\hat{\lambda}_{\mathtt{s}}^{(\mathtt{m})\star\mathtt{D}}\backsimeq\frac{q_{\mathtt{D,m}}P_{\mathtt{m}}^{\frac{2}{\alpha}}\delta\left(\gamma_{\mathtt{m}}^{\mathtt{D}},\alpha\right)}{q_{\mathtt{D,s}}\bigl(\hat{B}_{\mathtt{D,s}}^{(\mathtt{m})}\bigr)^{\frac{2}{\alpha}}P_{\mathtt{s}}^{\frac{2}{\alpha}}\delta\bigl(\gamma_{\mathtt{s}}^{\mathtt{D}},\alpha\bigr)}$,
and we derive that $\hat{\lambda}_{\mathtt{s}}^{(\mathtt{m})\star\mathtt{D}}$
is proportional to $q_{\mathtt{D,m}}$ and inversely proportional
to $q_{\mathtt{D,s}}$, (iii) for UL case, $\hat{\lambda}_{\mathtt{s}}^{(\mathtt{m})\star\mathtt{U}}$
increases with $q_{\mathtt{D,m}}$ while decreases with $q_{\mathtt{D,s}}$.

\subsection{Effects of bias factor}

\begin{figure}[t]
\centering \includegraphics[scale=0.35]{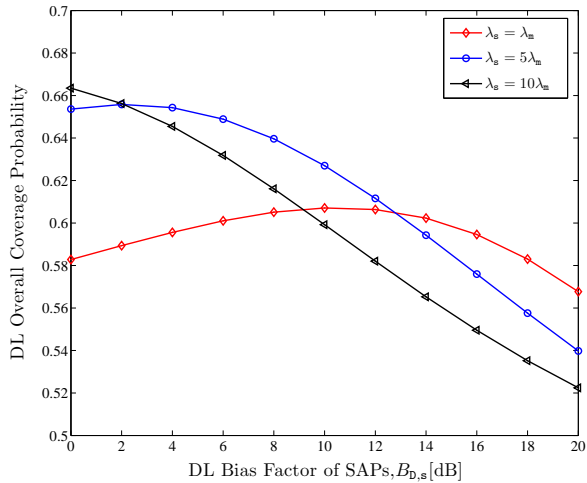}

\protect\caption{\label{fig:Cove-bias-fully-half}Overall downlink coverage probability
as a function of DL bias factor of SAPs $B_{\mathtt{D,s}}$, ($\lambda_{\mathtt{u}}=10^{3}\lambda_{\mathtt{m}}$,
$\{q_{\mathtt{D,m}},q_{\mathtt{D,s}}\}=\{0.5,0.5\}$, $\{B_{\mathtt{D,m}},B_{\mathtt{U,m}},B_{\mathtt{U,s}}\}=\{1,1,1\}$,
$\zeta=0.01$, $\rho_{\mathtt{s}}=\rho_{\mathtt{d}}=-60$ dBm).}
\end{figure}

Figure \ref{fig:Cove-bias-fully-half} depicts the DL coverage probability
as a function of the DL bias factor $B_{\mathtt{D,s}}$. We observe
that increasing the density of SAPs $\lambda_{\mathtt{s}}$ decreases
the optimal $B_{\mathtt{D,s}}$. This is due to the fact that a larger
$\lambda_{\mathtt{s}}$ inflicts more interference on the small cell
mobile users, and decreasing $B_{\mathtt{D,s}}$ helps to increase
the overall coverage probability by shifting small cell mobile users
with low SIR to the macro tier. It shows that with the analytical
framework, we can derive the optimal $\hat{B}_{\mathtt{D,s}}^{(\mathtt{m})}$
and $\hat{B}_{\mathtt{U,s}}^{(\mathtt{m})}$ that maximize the overall
coverage probability in DL and UL.

\emph{Optimization of bias factor}: In the fully-loaded network, by
taking the first-order derivative of $\bar{\mathbb{P}}_{\mathtt{D}}$
and $\bar{\mathbb{P}}_{\mathtt{U}}$ with respect to $\hat{B}_{\mathtt{D,s}}^{(\mathtt{m})}$
and $\hat{B}_{\mathtt{U,s}}^{(\mathtt{m})}$, we can derive the optimal
$\hat{B}_{\mathtt{D,s}}^{(\mathtt{m})\star}$ and $\hat{B}_{\mathtt{U,s}}^{(\mathtt{m})\star}$
for DL and UL transmissions as
\begin{equation}
\hat{B}_{\mathtt{D,s}}^{(\mathtt{m})\star}=\Bigl(\frac{P_{\mathtt{m}}^{\frac{2}{\alpha}}q_{\mathtt{D,m}}\delta(\gamma_{\mathtt{m}}^{\mathtt{D}},\alpha)+(1-q_{\mathtt{D,m}})\bigl(Q_{\mathtt{m}}\gamma_{\mathtt{m}}^{\mathtt{D}}\bigr)^{\frac{2}{\alpha}}}{\hat{\lambda}_{\mathtt{s}}^{(\mathtt{m})}\bigl(P_{\mathtt{s}}^{\frac{2}{\alpha}}q_{\mathtt{D,s}}\delta(\gamma_{\mathtt{s}}^{\mathtt{D}},\alpha)+(1-q_{\mathtt{D,s}})C(\alpha)\bigl(Q_{\mathtt{s}}\gamma_{\mathtt{s}}^{\mathtt{D}}\bigr)^{\frac{2}{\alpha}}\bigr)}\Bigr)^{\frac{\alpha}{2}}\label{eq:Op_bias_D}
\end{equation}
\begin{equation}
\hat{B}_{\mathtt{U,s}}^{(\mathtt{m})\star}=\Bigl(\frac{\bigl(\gamma_{\mathtt{m}}^{\mathtt{U}}\bigr)^{\frac{2}{\alpha}}\bigl(q_{\mathtt{D,m}}P_{\mathtt{m}}^{\frac{2}{\alpha}}+(1-q_{\mathtt{D,m}})Q_{\mathtt{m}}^{\frac{2}{\alpha}}\bigr)}{\hat{\lambda}_{\mathtt{s}}^{(\mathtt{m})}\bigl(\gamma_{\mathtt{s}}^{\mathtt{U}}\bigr)^{\frac{2}{\alpha}}\bigl(q_{\mathtt{D,s}}P_{\mathtt{s}}^{\frac{2}{\alpha}}+(1-q_{\mathtt{D,s}})Q_{\mathtt{s}}^{\frac{2}{\alpha}}\bigr)}\Bigr)^{\frac{\alpha}{2}}.\label{eq:Op_bias_U}
\end{equation}

\section{Network Access Design }

In this section, we first study the D2D enhanced network from a throughput
perspective and demonstrate the substantial throughput gain achieved
by D2D transmissions. Then we investigate the network access scheme
both from a coverage and throughput perspective. To emphasize the
benefit of network access scheme, we compare the proposed CSMA scheme
with the random access scheme ALOHA.

\begin{figure}[t]
\centering\includegraphics[scale=0.35]{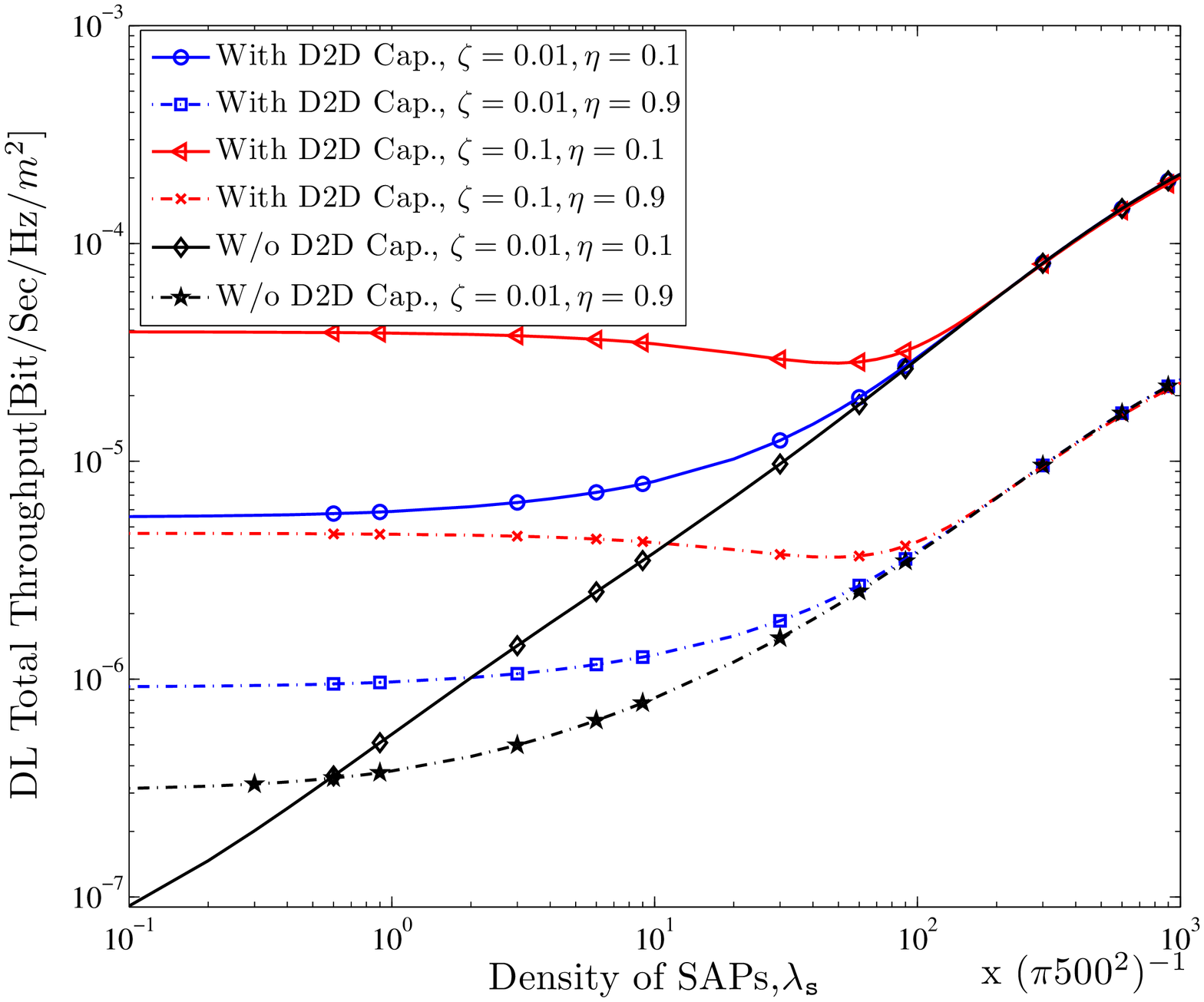}

\protect\caption{\label{fig:Thrpt_dens}Downlink total throughput vs. $\lambda_{\mathtt{s}}$
($\lambda_{\mathtt{u}}=10^{3}\lambda_{\mathtt{m}}$, $\{q_{\mathtt{D,m}},q_{\mathtt{D,s}}\}=\{0.5,0.5\}$,
$\{B_{\mathtt{D,m}},B_{\mathtt{D,s}},B_{\mathtt{U,m}},B_{\mathtt{U,s}}\}=\{1,1,1,1\}$,
$\rho_{\mathtt{s}}=\rho_{\mathtt{d}}=-60$ dBm). }
\end{figure}

Figure \ref{fig:Thrpt_dens} presents the DL network throughput with
and without D2D capabilities as a function of $\lambda_{\mathtt{s}}$
for different values of the bandwidth partition factor $\eta$. Without
D2D capabilities, the potential D2D transmitters can only associate
with the infrastructure in UL, and those potential D2D receivers are
blocked in the current timeslot. Comparing the curves with and without
D2D capabilities for the same $\zeta$, we observe that even a small
D2D user fraction ($\zeta=0.01$) results in a considerable throughput
gain. In the D2D enhanced network, we observe that allocating more
spectrum to the small cell tier leads to a larger network throughput,
ascribed to the high outage capacity of D2D users and the spatial
reuse gain from small cell transmissions. Figure \ref{fig:Thrpt_dens}
also illustrates that the network throughput benefits from a higher
fraction of potential D2D users $\zeta$. Interestingly, the network
throughput with $\zeta=0.1$ features a convex behavior as a function
of $\lambda_{\mathtt{s}}$. The initial decrease of network throughput
follows from the decline of the retaining probability with $\lambda_{\mathtt{s}}$
as indicated in \prettyref{eq:D2D_activity}. Compared with Fig. \ref{fig:Ave_Cove_dense},
we also observe that the D2D user fraction $\zeta$ leads to a tradeoff
between the overall coverage probability and the total network throughput
for fully-loaded network. A larger $\zeta$ improves the network throughput,
yet deteriorates the overall coverage probability due to the interference
inflicted on the small cell tier. The results presented in Fig. \ref{fig:Thrpt_dens}
indicate that the bandwidth partition strongly affects the network
throughput. In the following, we derive the optimal bandwidth partition
factor $\eta^{\star}$ by taking the DL network throughput presented
in \prettyref{eq:thrpt_D} as an example.

\emph{Optimization of bandwidth partition:} If $\mathcal{T}_{\mathtt{m}}^{\mathtt{D}}>\mathcal{T}_{\mathtt{s}}^{\mathtt{D}}+\frac{1}{2}\mathcal{T}_{\mathtt{d}}$,
$\mathcal{T}_{\mathtt{D}}(\eta;\rho_{\mathtt{s}};\rho_{\mathtt{d}})$
is a monotone increasing function of $\eta$, we have $\eta^{\star}=1$
and $\mathcal{T}_{\mathtt{D}}^{\star}(\eta^{\star};\rho_{\mathtt{s}};\rho_{\mathtt{d}})=\mathcal{T}_{\mathtt{m}}^{\mathtt{D}}$;
if $\mathcal{T}_{\mathtt{m}}^{\mathtt{D}}<\mathcal{T}_{\mathtt{s}}^{\mathtt{D}}+\frac{1}{2}\mathcal{T}_{\mathtt{d}},$
$\mathcal{T}_{\mathtt{D}}(\eta;\rho_{\mathtt{s}};\rho_{\mathtt{d}})$
monotonously decreases with $\eta$, thereby $\eta^{\star}=0$ and
$\mathcal{T}_{\mathtt{D}}^{\star}(\eta^{\star};\rho_{\mathtt{s}};\rho_{\mathtt{d}})=\mathcal{T}_{\mathtt{s}}^{\mathtt{D}}+\frac{1}{2}\mathcal{T}_{\mathtt{d}}$;
if $\mathcal{T}_{\mathtt{m}}^{\mathtt{D}}=\mathcal{T}_{\mathtt{s}}^{\mathtt{D}}+\frac{1}{2}\mathcal{T}_{\mathtt{d}}$,
$\mathcal{T}_{\mathtt{D}}(\eta;\rho_{\mathtt{s}};\rho_{\mathtt{d}})$
is a constant and does not change with $\eta$. The result is intuitive,
which means giving more bandwidth to the dominant tier is beneficial
to the total network throughput.

In this work, we use the distributed network access scheme CSMA to
control the channel access of D2D transmitters and protect the ongoing
small cell transmissions. We study the network access scheme both
from a coverage and throughput perspective. We refer to Fig. \ref{fig:D2D_Small_cove_rho_s_d},
which depicts that both the coverage probabilities of small cell tier
and D2D user deteriorate with $\rho_{\mathtt{s}}$. Similar effect
can be seen for $\rho_{\mathtt{d}}$. This is due to the fact that
the retaining probability and corresponding $\lambda_{\mathtt{d}}$
increase with $\rho_{\mathtt{s}}$ and $\rho_{\mathtt{d}}$. Thus,
in terms of the overall coverage probability for infrastructure based
transmissions and typical D2D user, the optimal sensing threshold
is given by $\rho_{\mathtt{s}}^{\star}=0$ and $\rho_{\mathtt{d}}^{\star}=0$.
However, the absence of D2D transmissions results in reduced network
throughput.

\begin{figure}[t]
\subfloat[]{\centering\includegraphics[scale=0.35]{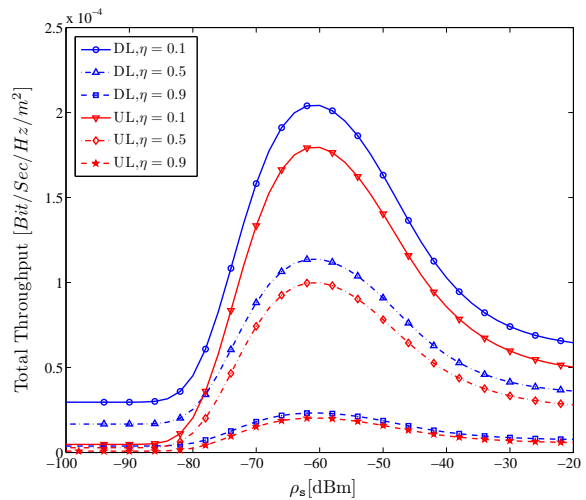}}

\subfloat[]{\centering\includegraphics[scale=0.35]{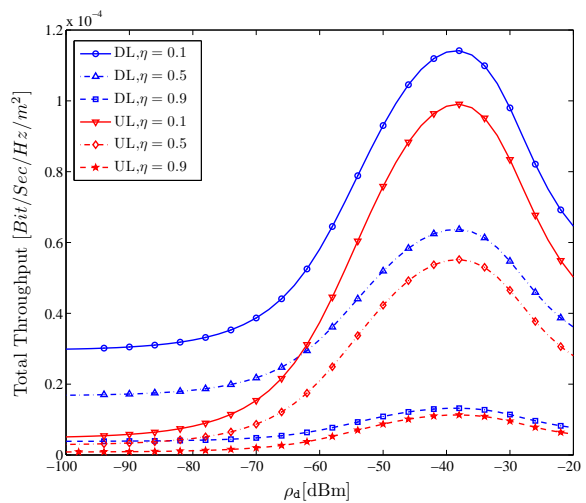}}\protect\caption{\label{fig:Thrpt_sense}Total network throughput vs. protection threshold
$\rho_{\mathtt{s}}$ or contention threshold $\rho_{\mathtt{d}}$,
(a) for $\rho_{\mathtt{s}}$, with $\rho_{\mathtt{d}}=-20$ dBm, (b)
for $\rho_{\mathtt{d}}$, with $\rho_{\mathtt{s}}=-20$ dBm, ($\lambda_{\mathtt{u}}=10^{4}\lambda_{\mathtt{m}}$,
$\lambda_{\mathtt{s}}=100\lambda_{\mathtt{m}}$, $\{q_{\mathtt{D,m}},q_{\mathtt{D,s}}\}=\{0.5,0.5\}$,
$\zeta=0.1$). }
\end{figure}

Figure \ref{fig:Thrpt_sense} depicts the total network throughput
as a function of $\rho_{\mathtt{s}}$ and $\rho_{\mathtt{d}}$. From
Fig. \ref{fig:Thrpt_sense}(a), we observe that the network throughput
exhibits a concave behavior with respect to $\rho_{\mathtt{s}}$.
This is caused by the opposite effects of $\rho_{\mathtt{s}}$ on
$\lambda_{\mathtt{d}}$ and the coverage probability of the small
cell tier and typical D2D user, a tradeoff between coverage probability
and D2D user activity that is made explicit in the expressions of
the network throughput \prettyref{eq:thrpt_D} and \prettyref{eq:thrpt_U}.
From Fig. \ref{fig:Thrpt_sense}(b), we notice a similar effect of
$\rho_{\mathtt{d}}$ on the network throughput. As for the coverage
analysis, the effect of $\rho_{\mathtt{s}}$ on the network throughput
is more evident than the effect of $\rho_{\mathtt{d}}$. This can
be understood by the effectiveness of the protection threshold $\rho_{\mathtt{s}}$
in controlling the mutual interference and improving the coverage
probability. In addition, we observe that the optimal $\rho_{\mathtt{d}}^{\star}$
is larger than $\rho_{\mathtt{s}}^{\star}$ due to the smaller effect
of $\rho_{\mathtt{d}}$ on the D2D retaining probability $\beta$.
Figure \ref{fig:Thrpt_sense} also shows that in the dense scenario
($\lambda_{\mathtt{s}}=100\lambda_{\mathtt{m}}$), giving more bandwidth
to the small cell tier can increase the total network throughput.
The presented results show that the proposed analytical framework
can be used to determine $\rho_{\mathtt{s}}^{\star}$ or $\rho_{\mathtt{d}}^{\star}$
that maximize the total network throughput.

\begin{figure}[t]
\subfloat[]{\centering\includegraphics[scale=0.35]{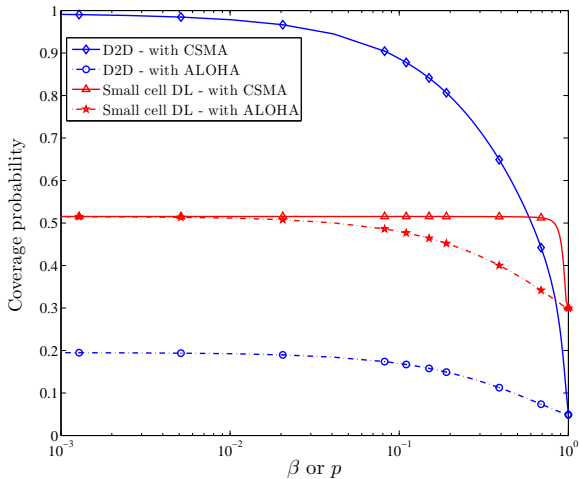}}

\subfloat[]{\centering\includegraphics[scale=0.35]{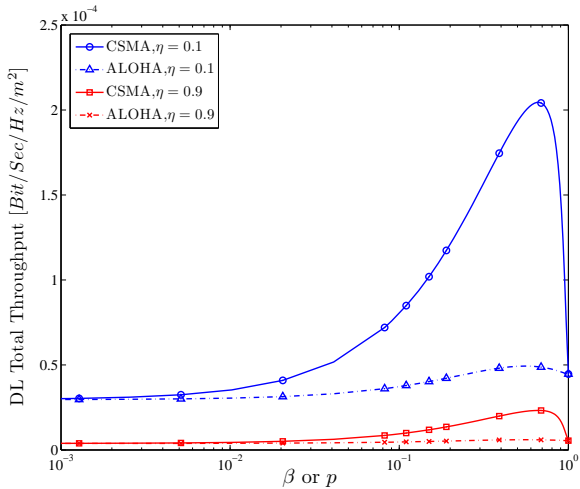}}

\protect\caption{\label{fig:CSMA_ALOHA_Compare}Comparison of coverage probability
and DL total network throughput from CSMA and ALOHA as a function
of retaining probability $\beta$ or access probability $p$, (a)
for coverage probability, (b) for downlink total network throughput,
($\lambda_{\mathtt{u}}=10^{4}\lambda_{\mathtt{m}}$, $\lambda_{\mathtt{s}}=100\lambda_{\mathtt{m}}$,
$\{q_{\mathtt{D,m}},q_{\mathtt{D,s}}\}=\{0.5,0.5\}$, $\zeta=0.1$,
$\rho_{\mathtt{d}}=-20$ dBm). }
\end{figure}

In Fig. \ref{fig:CSMA_ALOHA_Compare}, we depict the coverage probability
and DL network throughput as a function of retaining probability $\beta$
or access probability $p$ when D2D users utilize CSMA and ALOHA.
We keep $\rho_{\mathtt{d}}=-20$ dBm, and change $\rho_{\mathtt{s}}$
to alter $\beta$.%
\footnote{Note that with the given parameters, the protection threshold $\rho_{\mathtt{d}}=-20$
dBm leads to negligible contention among D2D transmitters such that
the network performance is dominated by $\rho_{\mathtt{s}}$.%
} With ALOHA, each D2D transmitter activates its transmission with
a certain probability $p$, $p\in(0,1]$, such that the active D2D
transmitters form a PPP with density $p\lambda_{\mathtt{d}}$. To
make a fair comparison, let $p=\beta$. From Fig. \ref{fig:CSMA_ALOHA_Compare}(a)
and Fig. \ref{fig:CSMA_ALOHA_Compare}(b), we observe that the proposed
CSMA scheme has prominent advantage over ALOHA in both coverage probability
and network throughput, which indicates that better network performance
can be achieved by a careful network access design.

\emph{Optimization of $\rho_{\mathtt{s}}$ and $\rho_{\mathtt{d}}$}:
The optimal sensing thresholds have been determined by using a numerical
search with limited computational complexity. We take the DL total
throughput derived in \prettyref{eq:thrpt_D} as an example. Since
$\mathcal{T}_{\mathtt{D}}(\eta;\rho_{\mathtt{s}};\rho_{\mathtt{d}})$
is a continuous function of $\rho_{\mathtt{s}}$ over {[}0,$\infty${]},
there exists at least one optimal $\rho_{\mathtt{s}}^{\star}$ where
$\mathcal{T}_{\mathtt{D}}(\eta;\rho_{\mathtt{s}};\rho_{\mathtt{d}})$
is maximized. Considering the signal attenuation characteristics,
$\rho_{\mathtt{s}}^{\star}$ is upper bounded by the transmit power
of D2D user $Q_{\mathtt{d}}$. Therefore, we have $\rho_{\mathtt{s}}^{\star}\in[0,Q_{\mathtt{d}}]$,
which reduces the complexity of the search process. Similarly, we
can get $\rho_{\mathtt{d}}^{\star}\in[0,Q_{\mathtt{d}}]$. The joint
optimization of $\rho_{\mathtt{s}}$ and $\rho_{\mathtt{d}}$ may
be implemented with a two-stage optimization method where we first
fix $\rho_{\mathtt{d}}$ and optimize the network throughput with
respect to $\rho_{\mathtt{s}}$, followed by the optimization with
respect to $\rho_{\mathtt{d}}$.

\section{Conclusion}

In this work, we studied a two-tier D2D enhanced HCN operating with
dynamic TDD, where the D2D transmitters follow a CSMA scheme. We proposed
a simple PPP model for the active D2D users and verified the accuracy
by extensive simulations. We presented an analytical framework to
evaluate the load-aware coverage probability and network throughput.
The proposed model allows us to analyze the non-trivial system behavior
of dynamic TDD networks and to quantify the effect of most important
network parameters such as the UL/DL configuration, base station density,
and bias factor on the coverage probability, and the bandwidth partition
on the total network throughput. We provided guidelines on the optimal
design of the network access scheme. Possible future directions to
extend this work are to include a dynamic traffic model in our framework
and consider the spatio-temporal correlations in the dynamic TDD network.

\appendix{}

\emph{A. Proof of Lemma 2}

Assuming $T_{0}$ locates at the origin and by Slivnyak's Theorem,
the point process of the D2D contenders forms a PPP. By using the
Palm distribution $P_{0}$, we derive the retaining probability $\beta$
of a potential D2D transmitter as follows:
\begin{eqnarray*}
\beta & = & P_{0}[U_{0}=1]\\
 & = & \mathbb{E}_{0}\Bigl[\prod_{Y_{j}\in\Phi_{\mathtt{s}}^{\mathtt{D}}}\mathbf{1}_{\bigl(\frac{Q_{\mathtt{d}}h_{oj}}{\Vert T_{0}-Y_{j}\Vert^{\alpha}}<\rho_{\mathtt{s}}\bigr)}\prod_{Z_{l}\in\Phi_{\mathtt{u},\mathtt{s}}^{\mathtt{T}}}\mathbf{1}_{\bigl(\frac{Q_{\mathtt{d}}h_{ol}}{\Vert T_{0}-Z_{l}\Vert^{\alpha}}<\rho_{\mathtt{s}}\bigr)}\\
 &  & \times\prod_{T_{k}\in\Phi_{\mathtt{d}}\backslash T_{0}}\bigl(\mathbf{1}_{(t_{0}\leq t_{k})}+\mathbf{1}_{(t_{0}>t_{k})}\mathbf{1}_{\bigl(\frac{Q_{\mathtt{d}}h_{ok}}{\Vert T_{0}-T_{k}\Vert^{\alpha}}<\rho_{\mathtt{d}}\bigr)}\bigr)\Bigr]\\
 & \overset{(a)}{=} & \mathbb{E}_{0}\Bigl[\prod_{Y_{j}\in\Phi_{\mathtt{s}}^{\mathtt{D}}}\bigl(1-e^{-\frac{\rho_{\mathtt{s}}}{Q_{\mathtt{d}}}\Vert Y_{j}\Vert^{\alpha}}\bigr)\Bigr]\mathbb{E}_{0}\Bigl[\prod_{Z_{l}\in\Phi_{\mathtt{u},\mathtt{s}}^{\mathtt{T}}}\bigl(1-e^{-\frac{\rho_{\mathtt{s}}}{Q_{\mathtt{d}}}\Vert Z_{l}\Vert^{\alpha}}\bigr)\Bigr]\\
 &  & \times\int_{0}^{1}\mathbb{E}_{0}\Bigl[\prod_{T_{k}\in\Phi_{\mathtt{d}}\backslash T_{0}}\mathbb{E}\bigl[\left(1-t\right)+t\bigl(1-e^{-\frac{\rho_{\mathtt{d}}}{Q_{\mathtt{d}}}\Vert T_{k}\Vert^{\alpha}}\bigr)\bigr]\Bigr]dt\\
 & \overset{(b)}{=} & \exp\bigl(-\lambda_{\mathtt{s}}^{\mathtt{D}}\int_{\mathbb{R}^{2}}e^{-\frac{\rho_{\mathtt{s}}}{Q_{\mathtt{d}}}\Vert x\Vert^{\alpha}}dx\bigr)\exp\bigl(-\lambda_{\mathtt{u,s}}^{\mathtt{T}}\int_{\mathbb{R}^{2}}e^{-\frac{\rho_{\mathtt{s}}}{Q_{\mathtt{d}}}\Vert x\Vert^{\alpha}}dx\bigr)\\
 &  & \times\int_{0}^{1}\exp\bigl(-\zeta\lambda_{\mathtt{u}}t\int_{\mathbb{R}^{2}}e^{-\frac{\rho_{\mathtt{d}}}{Q_{\mathtt{d}}}\Vert x\Vert^{\alpha}}dx\bigr)dt\\
 & \overset{\left(c\right)}{=} & \exp\bigl(-\bigl(\lambda_{\mathtt{s}}^{\mathtt{D}}+\lambda_{\mathtt{u,s}}^{\mathtt{T}}\bigr)\mathcal{K}_{o,\mathtt{s}}\bigr)\frac{1-\exp\bigl(-\zeta\lambda_{\mathtt{u}}\mathcal{K}_{o,\mathtt{d}}\bigr)}{\zeta\lambda_{\mathtt{u}}\mathcal{K}_{o,\mathtt{d}}},
\end{eqnarray*}
where $\mathcal{K}_{o,\mathtt{s}}=\frac{2\pi\Gamma\left(\frac{2}{\alpha}\right)}{\alpha\bigl(\frac{\rho_{\mathtt{s}}}{Q_{\mathtt{d}}}\bigr)^{\frac{2}{\alpha}}}$
and $\mathcal{K}_{o,\mathtt{d}}=\frac{2\pi\Gamma\left(\frac{2}{\alpha}\right)}{\alpha\bigl(\frac{\rho_{\mathtt{d}}}{Q_{\mathtt{d}}}\bigr)^{\frac{2}{\alpha}}}$,
(a) follows by the independence of PPPs and the expectation over the
channel gains $h$, (b) is obtained by using the probability generating
functional (PGFL) of the PPP \cite{SGAI} of $\Phi_{\mathtt{s}}^{\mathtt{D}}$,
$\Phi_{\mathtt{u,s}}^{\mathtt{T}}$ and $\Phi_{\mathtt{d}}$, and
(c) evaluates the given integrals by changing to polar coordinates
and the use of Gamma function.

\emph{B. Proof of Theorem 1}

First, we derive the  coverage probability of the macro tier as follows.
For DL mode,
\begin{eqnarray}
 & \mathbb{P}_{\mathtt{m}}^{\mathtt{D}}\nonumber \\
 & \overset{(a)}{=} & \int_{0}^{\infty}\mathbb{E}_{\Phi_{\mathtt{m}}^{\mathtt{D}},\Phi_{\mathtt{u},\mathtt{m}}^{\mathtt{T}}}\Bigl[\exp\bigl(-\frac{\gamma_{\mathtt{m}}^{\mathtt{D}}r^{\alpha}}{P_{\mathtt{m}}}\bigl(I_{\mathtt{D\rightarrow D}}^{\mathtt{(m)}}+I_{\mathtt{U\rightarrow D}}^{\mathtt{(m)}}\bigr)\bigr)\Bigr]f_{Y_{\mathtt{D,m}}}\left(r\right)dr\nonumber \\
 & \overset{(b)}{=} & \int_{0}^{\infty}\mathbb{E}_{\Phi_{\mathtt{m}}^{\mathtt{D}}}\Bigl[\exp\Bigl(-\frac{\gamma_{\mathtt{m}}^{\mathtt{D}}r^{\alpha}}{P_{\mathtt{m}}}I_{\mathtt{D\rightarrow D}}^{\mathtt{(m)}}\Bigr)\Bigr]\nonumber \\
 &  & \times\mathbb{E}_{\Phi_{\mathtt{u,m}}^{\mathtt{T}}}\Bigl[\exp\Bigl(-\frac{\gamma_{\mathtt{m}}^{\mathtt{D}}r^{\alpha}}{P_{\mathtt{m}}}I_{\mathtt{U\rightarrow D}}^{\mathtt{(m)}}\Bigr)\Bigr]f_{Y_{\mathtt{D,m}}}\left(r\right)dr\nonumber \\
 & \overset{(c)}{=} & \int_{0}^{\infty}\exp\left(-\pi r^{2}\lambda_{\mathtt{m}}^{\mathtt{D}}\delta\left(\gamma_{\mathtt{m}}^{\mathtt{D}},\alpha\right)\right)\nonumber \\
 &  & \times\exp\biggl(-\pi r^{2}\lambda_{\mathtt{u,m}}^{\mathtt{T}}C\left(\alpha\right)\Bigl(\frac{Q_{\mathtt{m}}}{P_{\mathtt{m}}}\gamma_{\mathtt{m}}^{\mathtt{D}}\Bigr)^{^{\frac{2}{\alpha}}}\biggr)f_{Y_{\mathtt{D,m}}}\left(r\right)dr\nonumber \\
 & \overset{(d)}{=} & \frac{q_{\mathtt{D,m}}\lambda_{\mathtt{m}}}{\lambda_{\mathtt{m}}^{\mathtt{D}}\mathcal{A}_{\mathtt{D}\mathtt{,m}}\delta\left(\gamma_{\mathtt{m}}^{\mathtt{D}},\alpha\right)+\lambda_{\mathtt{u,m}}^{\mathtt{T}}\mathcal{A}_{\mathtt{D}\mathtt{,m}}C\left(\alpha\right)\bigl(\frac{Q_{\mathtt{m}}\gamma_{\mathtt{m}}^{\mathtt{D}}}{P_{\mathtt{m}}}\bigr)^{\frac{2}{\alpha}}+q_{\mathtt{D,m}}\lambda_{\mathtt{m}}},\nonumber \\
\label{eq:Cov_dm}
\end{eqnarray}
where $C\left(\alpha\right)=\frac{2\pi/\alpha}{\sin\left(2\pi/\alpha\right)}$,
and $\delta\left(\beta,\alpha\right)=\int_{\beta^{-\frac{2}{\alpha}}}^{\infty}\frac{\beta^{\frac{2}{\alpha}}}{1+u^{\frac{\alpha}{2}}}du$,
(a) follows by taking expectation over the channel gains $h$, (b)
is due to the independence of the PPPs, (c) results from the Laplace
transform, where the first exponential term is caused by the fact
that the active interfering MBSs can not stay within the disk $b\left(0,r\right)$.
Finally, (d) follows by integrating with respect to the PDF $f_{Y_{\mathtt{D},\mathtt{m}}}(y)$
as defined in \prettyref{eq:Dis_PDF_DL}. For UL mode, the  coverage
probability is given by
\begin{eqnarray}
 & \mathbb{P}_{\mathtt{m}}^{\mathtt{U}}\nonumber \\
 & = & \int_{0}^{\infty}\mathbb{E}_{\Phi_{\mathtt{m}}^{\mathtt{D}},\Phi_{\mathtt{u},\mathtt{m}}^{\mathtt{T}}}\Bigl[\exp\bigl(-\frac{\gamma_{\mathtt{m}}^{\mathtt{U}}r^{\alpha}}{Q_{\mathtt{m}}}\bigl(\ensuremath{I_{\mathtt{D\rightarrow U}}^{\mathtt{(m)}}+I_{\mathtt{U\rightarrow U}}^{(\mathtt{m})}}\bigr)\bigr)\Bigr]f_{Y_{\mathtt{U},\mathtt{m}}}\left(r\right)dr\nonumber \\
 & = & \int_{0}^{\infty}\exp\bigl(-\pi r^{2}\lambda_{\mathtt{m}}^{\mathtt{D}}C\left(\alpha\right)\bigl(\frac{P_{\mathtt{m}}}{Q_{\mathtt{m}}}\gamma_{\mathtt{m}}^{\mathtt{U}}\bigr)^{\frac{2}{\alpha}}\bigr)\nonumber \\
 &  & \times\exp\bigl(-\pi r^{2}\lambda_{\mathtt{u,m}}^{\mathtt{T}}C\left(\alpha\right)\left(\gamma_{\mathtt{m}}^{\mathtt{U}}\right)^{\frac{2}{\alpha}}\bigr)f_{Y_{\mathtt{U},\mathtt{m}}}\left(r\right)dr\nonumber \\
 & = & \frac{(1-q_{\mathtt{D,m}})\lambda_{\mathtt{m}}}{C\left(\alpha\right)\bigl(\gamma_{\mathtt{m}}^{\mathtt{U}}\bigr)^{\frac{2}{\alpha}}\mathcal{A}_{\mathtt{U}\mathtt{,m}}\bigl(\lambda_{\mathtt{m}}^{\mathtt{D}}\bigl(\frac{P_{\mathtt{m}}}{Q_{\mathtt{m}}}\bigr)^{\frac{2}{\alpha}}+\lambda_{\mathtt{u,m}}^{\mathtt{T}}\bigr)+(1-q_{\mathtt{D,m}})\lambda_{\mathtt{m}}}.\nonumber \\
\label{eq:Cov_um}
\end{eqnarray}

With regard to the small cell tier, the Laplace transform of the aggregate
interference from active SAPs and transmitting mobile users can be
derived similar to \prettyref{eq:Cov_dm} and \prettyref{eq:Cov_um},
respectively. In the following, we focus on the Laplace transform
of the aggregate interference from active D2D transmitters. As is
shown in Fig. \ref{fig:D2D_small cell}, we have
\begin{eqnarray}
\mathcal{L}_{I}(s) & = & \mathbb{E}_{\Phi_{\mathtt{d}}}\Biggl[\exp\biggl(-s\Bigl(\sum_{z\in\Phi_{\mathtt{d}}\cap\overline{\mathcal{H}}}Q_{\mathtt{d}}h_{oz}z^{-\alpha}\Bigr)\biggr)\Biggr]\nonumber \\
 & \times & \mathbb{E}_{\Phi_{\mathtt{d}}}\Biggl[\exp\biggl(-s\Bigl(\sum_{z\in\Phi_{\mathtt{d}}\cap\mathcal{H}\cap\overline{\mathcal{B}}}Q_{\mathtt{d}}h_{oz}z^{-\alpha}\Bigr)\biggr)\Biggr]\label{eq:50}
\end{eqnarray}
where the first term is related to the interferer distributed outside
the big circle of radius $\Vert y_{0}\Vert+\iota_{\mathtt{s}}$, i.e.
the shaded region $\Phi_{\mathtt{d}}\cap\overline{\mathcal{H}}$ ,
and can be easily derived. The second term corresponds to the interferer
scattered over the shaded region within the big circle, i.e. $\Phi_{\mathtt{d}}\cap\mathcal{H}\cap\overline{\mathcal{B}}$.
Two cases are differentiated for the calculation of the Laplace transform:
$\Vert y_{0}\Vert\leq\iota_{\mathtt{s}}$ (see Fig. \ref{fig:D2D_small cell}(a))
and $\Vert y_{0}\Vert>\iota_{\mathtt{s}}$ (see Fig. \ref{fig:D2D_small cell}(b)).
Denote the Laplace transform of the interference from active D2D transmitters
in $\Phi_{\mathtt{d}}\cap\overline{\mathcal{H}}$ and $\Phi_{\mathtt{d}}\cap\mathcal{H}\cap\overline{\mathcal{B}}$
as $\mathcal{L}_{I_{out}}(s)$ and $\mathcal{L}_{I_{in}}(s)$, respectively.
Conditioned on the typical small cell link length being $\Vert y_{0}\Vert$,
we have
\begin{eqnarray}
 &  & \mathcal{L}_{I_{out}}(s\mid\Vert y_{0}\Vert)\nonumber \\
 & = & \mathbb{E}_{\Phi_{\mathtt{d}}}\Biggl[\exp\biggl(-s\Bigl(\sum_{z\in\Phi_{\mathtt{d}}\cap\overline{\mathcal{H}}}Q_{\mathtt{d}}h_{oz}z^{-\alpha}\Bigr)\biggr)\mid\Vert y_{0}\Vert\Biggr]\nonumber \\
 & \overset{(a)}{=} & \exp\Bigl(-2\pi\lambda_{\mathtt{d}}\int_{\iota_{\mathit{\mathtt{s}}+\Vert y_{0}\Vert}}^{\infty}\frac{y}{1+\frac{y^{\alpha}}{sQ_{\mathtt{d}}}}dy\Bigr)\nonumber \\
 & \overset{(b)}{=} & \exp\Bigl(-\pi\lambda_{\mathtt{d}}\bigl(\iota_{\mathtt{s}}+\Vert y_{0}\Vert\bigr)^{2}\delta\Bigl(\frac{sQ_{\mathtt{d}}}{\bigl(\iota_{\mathtt{s}}+\Vert y_{0}\Vert\bigr)^{\alpha}},\alpha\Bigr)\Bigr)\label{eq:I_out}
\end{eqnarray}
where (a) follows from the PGFL of the PPP, and (b) is due to $\delta\left(\beta,\alpha\right)=\int_{\beta^{-\frac{2}{\alpha}}}^{\infty}\frac{\beta^{\frac{2}{\alpha}}}{1+u^{\frac{2}{\alpha}}}du$,
with $\iota_{\mathtt{s}}+\Vert y_{0}\Vert$ denoting the distance
from the nearest interferer staying within $\Phi_{\mathtt{d}}\cap\overline{\mathcal{H}}$.

In the following, we focus on the computation of $\mathcal{L}_{I_{in}}(s)$.
For notational simplicity, we first define a $\mathcal{Z}$ function
as
\begin{equation}
\mathcal{Z}_{\theta_{l},\kappa_{l}}^{\theta_{u},\kappa_{u}}\left(s;Q\right)=\left(sQ\right)^{\frac{2}{\alpha}}\int_{\theta_{l}}^{\theta_{u}}\int_{\frac{\kappa_{l}^{2}}{(sQ)^{\frac{2}{\alpha}}}}^{\frac{\kappa_{u}^{2}}{\left(sQ\right)^{\frac{2}{\alpha}}}}\frac{1}{1+u^{\frac{2}{\alpha}}}dud\theta,\label{eq:Z_func}
\end{equation}
where $Q$ represents the transmit power of interferer, $\theta_{l}$,
$\theta_{u}$, and $\kappa_{l}$, $\kappa_{u}$ denote the lower bound
and upper bound of the angle and distance from the interferer distributed
in the shaded region $\Phi_{\mathtt{d}}\cap\mathcal{H}\cap\overline{\mathcal{B}}$.

For $\Vert y_{0}\Vert\leq\iota_{\mathtt{s}}$, denote $l_{OA}=||OA||$,
we have $l_{OA}=\sqrt{\iota_{\mathtt{s}}^{2}-\bigl(\Vert y_{0}\Vert\textrm{sin}\theta\bigr)^{2}}+\Vert y_{0}\Vert\cos\theta,\;\theta\in[0,\pi].$
The Laplace transform $\mathcal{L}_{I_{in}}(s\mid\Vert y_{0}\Vert\leq\iota_{\mathtt{s}})$
is derived as
\begin{eqnarray}
 &  & \mathcal{L}_{I_{in}}(s\mid\Vert y_{0}\Vert\leq\iota_{\mathtt{s}})\nonumber \\
 & = & \mathbb{E}_{\Phi_{\mathtt{d}}}\Bigl[\exp\bigl(-s(\sum_{z\in\Phi_{\mathtt{d}}\cap\mathcal{H}\cap\overline{\mathcal{B}}}Q_{\mathtt{d}}h_{oz}z^{-\alpha})\bigr)\mid\Vert y_{0}\Vert\leq\iota_{\mathtt{s}}\Bigr]\nonumber \\
 & \overset{(a)}{=} & \exp\bigl(-2\lambda_{\mathtt{d}}\int_{0}^{\pi}\int_{l_{OA}}^{\iota_{\mathtt{s}}+\Vert y_{0}\Vert}\frac{y}{1+\frac{y^{\alpha}}{sQ_{\mathtt{d}}}}dyd\theta\bigr)\nonumber \\
 & \overset{(b)}{=} & \exp\Bigl(-\lambda_{\mathtt{d}}\mathcal{Z}_{0,l_{OA}}^{\pi,\iota_{s}+\Vert y_{0}\Vert}\left(s;Q_{\mathtt{d}}\right)\Bigr),\label{eq:I_in}
\end{eqnarray}
where (a) follows by the PGFL of the PPP and by converting to polar
coordinates, and (b) follows by substituting the corresponding bounds
of the angle and distance into \prettyref{eq:Z_func}. By combining
\prettyref{eq:I_out} with \prettyref{eq:I_in}, we have
\begin{equation}
\mathcal{L}_{I}(s\mid\Vert y_{0}\Vert\leq\iota_{\mathtt{s}})=\mathcal{L}_{I_{out}}(s\mid\Vert y_{0}\Vert)\mathcal{L}_{I_{in}}(s\mid\Vert y_{0}\Vert\leq\iota_{\mathtt{s}}).
\end{equation}

For $\Vert y_{0}\Vert>\iota_{\mathtt{s}}$, denote $\Theta=\arcsin\bigl(\frac{\iota_{\mathtt{s}}}{\Vert y_{0}\Vert}\bigr)$,
$l_{OD}=\Vert OD\Vert$ and $l_{OC}=||OC||$, where $l_{OD}$ and
$l_{OC}$ can be found by simple geometric formulas, similar to $l_{OA}$
given before. The Laplace transform $\mathcal{L}_{I_{in}}(s\mid\Vert y_{0}\Vert>\iota_{\mathtt{s}})$
is given by
\begin{eqnarray}
 &  & \mathcal{L}_{I_{in}}(s\mid\Vert y_{0}\Vert>\iota_{\mathtt{s}})\nonumber \\
 & \overset{(a)}{=} & \exp\Bigl(-2\lambda_{\mathtt{d}}\bigl(\int_{0}^{\Theta}\int_{0}^{l_{OD}}\frac{y}{1+\frac{y^{\alpha}}{sQ_{\mathtt{d}}}}dyd\theta+\int_{0}^{\Theta}\int_{l_{OC}}^{\iota_{\mathtt{s}}+\Vert y_{0}\Vert}\nonumber \\
 &  & \times\frac{y}{1+\frac{y^{\alpha}}{sQ_{\mathtt{d}}}}dyd\theta+\int_{\Theta}^{\pi}\int_{0}^{\iota_{\mathtt{s}}+\Vert y_{0}\Vert}\frac{y}{1+\frac{y^{\alpha}}{sQ_{\mathtt{d}}}}dyd\theta\bigr)\Bigr)\nonumber \\
 & = & \exp\Bigl(-\lambda_{\mathtt{d}}\bigl(\mathcal{Z}_{0,0}^{\Theta,l_{OD}}\left(s;Q_{\mathtt{d}}\right)\nonumber \\
 &  & +\mathcal{Z}_{0,l_{OC}}^{\Theta,\iota_{\mathtt{s}}+\Vert y_{0}\Vert}\left(s;Q_{\mathtt{d}}\right)+\mathcal{Z}_{\Theta,0}^{\pi,\iota_{\mathtt{s}}+\Vert y_{0}\Vert}\left(s;Q_{\mathtt{d}}\right)\bigr)\Bigr),\label{eq:I_in_1}
\end{eqnarray}
where (a) is due to the fact that when $\theta<\arcsin\bigl(\frac{\iota_{\mathtt{s}}}{\Vert y_{0}\Vert}\bigr)$,
the line $\overline{OB}$ shown in Fig. 2(b) passes through the whole
exclusion region. By combining \prettyref{eq:I_out} with \prettyref{eq:I_in_1},
we obtain
\begin{equation}
\mathcal{L}_{I}(s\mid\Vert y_{0}\Vert>\iota_{\mathtt{s}})=\mathcal{L}_{I_{out}}(s\mid\Vert y_{0}\Vert)\mathcal{L}_{I_{in}}(s\mid\Vert y_{0}\Vert>\iota_{\mathtt{s}}).
\end{equation}

For DL transmission, substituting $I=I_{\mathtt{d\rightarrow D}}=\sum_{z\in\Phi_{\mathtt{d}}\backslash b\left(y_{0},\iota_{\mathtt{s}}\right)}Q_{\mathtt{d}}h_{oz}z^{-\alpha}$,
we derive
\begin{eqnarray*}
\mathbb{P}_{\mathtt{s}}^{\mathtt{D}} & = & \int_{0}^{\infty}\mathbb{E}_{\Phi_{\mathtt{s}}^{\mathtt{D}},\Phi_{\mathtt{u},\mathtt{s}}^{\mathtt{T}},\Phi_{\mathtt{d}}}\Bigl[\exp\Bigl(-\frac{\gamma_{\mathtt{s}}^{\mathtt{D}}r^{\alpha}}{P_{\mathtt{s}}}\bigl(I_{\mathtt{D\rightarrow D}}^{\mathtt{(s)}}\\
 &  & +I_{\mathtt{U\rightarrow D}}^{\mathtt{(s)}}+I_{\mathtt{d\rightarrow D}}\bigr)\Bigr)\Bigr]f_{Y_{\mathtt{D,s}}}\left(r\right)dr\\
 & \overset{(a)}{=} & \frac{\pi q_{\mathtt{D,s}}\lambda_{\mathtt{s}}}{\mathcal{A}_{\mathtt{D,s}}}\biggl[\int_{0}^{\iota_{\mathtt{s}}^{2}}e^{-\pi v\mathcal{F}}\mathcal{L}_{I_{\mathtt{d\rightarrow D}}}(\frac{\gamma_{\mathtt{s}}^{\mathtt{D}}v^{\frac{\alpha}{2}}}{P_{\mathtt{s}}}\mid v\leq\iota_{\mathtt{s}}^{2})dv\\
 &  & +\int_{\iota_{\mathtt{s}}^{2}}^{\infty}e^{-\pi v\mathcal{F}}\mathcal{L}_{I_{\mathtt{d\rightarrow D}}}(\frac{\gamma_{\mathtt{s}}^{\mathtt{D}}v^{\frac{\alpha}{2}}}{P_{\mathtt{s}}}\mid v>\iota_{\mathtt{s}}^{2})dv\biggr],
\end{eqnarray*}
where (a) follows by inserting $s=\frac{\gamma_{\mathtt{s}}^{\mathtt{D}}r^{\alpha}}{P_{\mathtt{s}}}$
and the change of variable $r^{2}\rightarrow v$, and
\begin{equation}
\mathcal{F\triangleq}\lambda_{\mathtt{s}}^{\mathtt{D}}\delta\bigl(\gamma_{\mathtt{s}}^{\mathtt{D}},\alpha\bigr)+\lambda_{\mathtt{u,s}}^{\mathtt{T}}C\left(\alpha\right)\bigl(\frac{Q_{\mathtt{s}}}{P_{\mathtt{s}}}\gamma_{\mathtt{s}}^{\mathtt{D}}\bigr)^{\frac{2}{\alpha}}+\frac{q_{\mathtt{D,s}}\lambda_{\mathtt{s}}}{\mathcal{A}_{\mathtt{D,s}}},\label{eq:F}
\end{equation}
where the first term and second term, respectively, relates to the
interference from DL SAPs and transmitting mobile users, and the last
term comes from $f_{Y_{\mathtt{D,s}}}\left(r\right)$ derived in \prettyref{eq:Dis_PDF_DL}.

For UL transmission, by substituting $I=I_{\mathtt{d\rightarrow U}}=\sum_{z\in\Phi_{\mathtt{d}}\backslash b\left(x_{0},\iota_{\mathtt{s}}\right)}Q_{\mathtt{d}}h_{oz}z^{-\alpha},$
we have
\begin{eqnarray*}
\mathbb{P}_{\mathtt{s}}^{\mathtt{U}} & = & \int_{0}^{\infty}\mathbb{E}_{\Phi_{\mathtt{s}}^{\mathtt{D}},\Phi_{\mathtt{u},\mathtt{s}}^{\mathtt{T}},\Phi_{\mathtt{d}}}\Bigl[\exp\Bigl(-\frac{\gamma_{\mathtt{s}}^{\mathtt{U}}r^{\alpha}}{Q_{\mathtt{s}}}\bigl(I_{\mathtt{D\rightarrow U}}^{\mathtt{(s)}}\\
 &  & +I_{\mathtt{U\rightarrow U}}^{\mathtt{(s)}}+I_{\mathtt{d\rightarrow U}}\bigr)\Bigr)\Bigr]f_{Y_{\mathtt{U,s}}}\left(r\right)dr\\
 & \overset{(a)}{=} & \frac{\pi\left(1-q_{\mathtt{D,s}}\right)\lambda_{\mathtt{s}}}{\mathcal{A}_{\mathtt{U,s}}}\biggl[\int_{0}^{\iota_{\mathtt{s}}^{2}}e^{-\pi v\mathcal{G}}\mathcal{L}_{I_{\mathtt{d\rightarrow U}}}(\frac{\gamma_{\mathtt{s}}^{\mathtt{U}}v^{\frac{\alpha}{2}}}{Q_{\mathtt{s}}}\mid v\leq\iota_{\mathtt{s}}^{2})dv\\
 &  & +\int_{\iota_{\mathtt{s}}^{2}}^{\infty}e^{-\pi v\mathcal{G}}\mathcal{L}_{I_{\mathtt{d\rightarrow U}}}(\frac{\gamma_{\mathtt{s}}^{\mathtt{U}}v^{\frac{\alpha}{2}}}{Q_{\mathtt{s}}}\mid v>\iota_{\mathtt{s}}^{2})dv\biggr],
\end{eqnarray*}
where (a) follows by inserting $s=\frac{\gamma_{\mathtt{s}}^{\mathtt{U}}r^{\alpha}}{Q_{\mathtt{s}}}$
and the change of variable $r^{2}\rightarrow v$, and
\begin{equation}
\mathcal{G\triangleq}C\left(\alpha\right)\bigl(\gamma_{\mathtt{s}}^{\mathtt{U}}\bigr)^{\frac{2}{\alpha}}\Bigl(\lambda_{\mathtt{s}}^{\mathtt{D}}\bigl(\frac{P_{\mathtt{s}}}{Q_{\mathtt{s}}}\bigr)^{\frac{2}{\alpha}}+\lambda_{\mathtt{u,s}}^{\mathtt{T}}\Bigr)+\frac{\left(1-q_{\mathtt{D,s}}\right)\lambda_{\mathtt{s}}}{\mathcal{A}_{\mathtt{U,s}}}.\label{eq:G}
\end{equation}
where the first term and second term, respectively, relates to the
interference from DL SAPs and transmitting mobile users, and the last
term comes from $f_{Y_{\mathtt{U,s}}}\left(r\right)$ derived in \prettyref{eq:Dis_PDF_UL}.
Furthermore, combining \prettyref{eq:Asso_D_U} with \prettyref{eq:Cov_dm},
\prettyref{eq:Cov_um}, $\mathbb{P}_{\mathtt{s}}^{\mathtt{D}}$ and
$\mathbb{P}_{\mathtt{s}}^{\mathtt{U}}$, we obtain the overall load-aware
coverage probabilities in DL and UL in \prettyref{eq:Ave_coverage}.

At last, we obtain the D2D receiver coverage probability. Due to the
small value of $r_{\mathtt{d}}$, we reasonably assume $r_{\mathtt{d}}\leq\iota_{\mathtt{s}}$
and $r_{\mathtt{d}}\leq\iota_{\mathtt{d}}$ to simplify the analysis.
Due to the CSMA scheme, it is equivalent to draw two exclusion regions
around the serving D2D transmitter with radius $\iota_{\mathtt{s}}$
and $\iota_{\mathtt{d}}$, within which, the small cell transmitters
and D2D transmitters are not allowed to exist, respectively. The typical
D2D receiver is assumed to be at the origin and the serving D2D transmitter
locates at $z_{0}$. The derivation of the coverage probability of
typical D2D receiver is similar to the coverage probability of small
cell tier with the case $\Vert y_{0}\Vert\leq\iota_{\mathtt{s}}$,
as shown in Fig. \ref{fig:D2D_small cell}(a). Thus, we skip the details
and give the results directly.
\begin{eqnarray}
\mathbb{P}_{\mathtt{d}} & = & \mathbb{E}_{\Phi_{\mathtt{s}}^{\mathtt{D}},\Phi_{\mathtt{u},\mathtt{s}}^{\mathtt{T}},\Phi_{\mathtt{d}}}\Bigl[\exp\Bigl(-\frac{\gamma_{\mathtt{d}}r_{\mathtt{d}}^{\alpha}}{Q_{\mathtt{d}}}\bigl(I_{\mathtt{D\rightarrow d}}^{\mathtt{(s)}}+I_{\mathtt{U\rightarrow d}}^{\mathtt{(s)}}+I_{\mathtt{d\rightarrow d}}\bigr)\Bigr)\Bigr]\nonumber \\
 & = & \exp\biggl(-\mathcal{I}_{1}\Bigl(\frac{\gamma_{\mathtt{d}}r_{\mathtt{d}}^{\alpha}}{Q_{\mathtt{d}}};P_{\mathtt{s}}\Bigr)\nonumber \\
 &  & -\mathcal{I}_{2}\Bigl(\frac{\gamma_{\mathtt{d}}r_{\mathtt{d}}^{\alpha}}{Q_{\mathtt{d}}};Q_{\mathtt{s}}\Bigr)-\mathcal{I}_{3}\Bigl(\frac{\gamma_{\mathtt{d}}r_{\mathtt{d}}^{\alpha}}{Q_{\mathtt{d}}};Q_{\mathtt{d}}\Bigr)\biggr),
\end{eqnarray}
where $\mathcal{I}_{1}\left(s;P_{\mathtt{s}}\right)$, $\mathcal{I}_{2}\left(s;Q_{\mathtt{s}}\right)$
and $\mathcal{I}_{3}\left(s;Q_{\mathtt{d}}\right)$ correspond to
the interference incurred by the DL SAPs, transmitting mobile users
and active D2D transmitters, and are given by \prettyref{eq:I_1},
\prettyref{eq:I_2} and \prettyref{eq:I_3}, respectively.

\bibliographystyle{IEEEtran}
\bibliography{TDD_D2D}

\begin{thebibliography}{10}
\providecommand{\url}[1]{#1}
\csname url@samestyle\endcsname
\providecommand{\newblock}{\relax}
\providecommand{\bibinfo}[2]{#2}
\providecommand{\BIBentrySTDinterwordspacing}{\spaceskip=0pt\relax}
\providecommand{\BIBentryALTinterwordstretchfactor}{4}
\providecommand{\BIBentryALTinterwordspacing}{\spaceskip=\fontdimen2\font plus
\BIBentryALTinterwordstretchfactor\fontdimen3\font minus
  \fontdimen4\font\relax}
\providecommand{\BIBforeignlanguage}[2]{{%
\expandafter\ifx\csname l@#1\endcsname\relax
\typeout{** WARNING: IEEEtran.bst: No hyphenation pattern has been}%
\typeout{** loaded for the language `#1'. Using the pattern for}%
\typeout{** the default language instead.}%
\else
\language=\csname l@#1\endcsname
\fi
#2}}
\providecommand{\BIBdecl}{\relax}
\BIBdecl

\bibitem{SCND}
T.~Q.~S. Quek, G.~de~la Roche, I.~Gven, and M.~Kountouris, \emph{{Small Cell
  Networks: Deployment, PHY Techniques, and Resource Management}}.\hskip 1em
  plus 0.5em minus 0.4em\relax New York, NY, USA: Cambridge Univ. Press, 2013.

\bibitem{TTHW}
H.~ElSawy and E.~Hossain, ``Two-tier hetnets with cognitive femtocells:
  Downlink performance modeling and analysis in a multichannel environment,''
  \emph{IEEE Trans. Mob. Comput.}, vol.~13, no.~3, pp. 649--663, March 2014.

\bibitem{OTCOD}
J.~Wen, M.~Sheng, X.~Wang, J.~Li, and H.~Sun, ``On the capacity of downlink
  multi-hop heterogeneous cellular networks,'' \emph{IEEE Trans. Wireless
  Commun.}, vol.~13, no.~8, pp. 4092--4103, Aug. 2014.

\bibitem{CSCN}
M.~Wildemeersch, T.~Q.~S. Quek, C.~Slump, and A.~Rabbachin, ``Cognitive small
  cell networks: Energy efficiency and trade-offs,'' \emph{IEEE Trans.
  Commun.}, vol.~61, no.~9, pp. 4016--4029, Sept. 2013.

\bibitem{DDCI}
D.~Feng, L.~Lu, Y.~Yuan-Wu, G.~Ye~Li, S.~Li, and G.~Feng, ``Device-to-device
  communications in cellular networks,'' \emph{IEEE Commun. Mag.}, vol.~52,
  no.~4, pp. 49--55, Apr. 2014.

\bibitem{GRAM_Song}
L.~Song, D.~Niyato, Z.~Han, and E.~Hossain, ``Game-theoretic resource
  allocation methods for device-to-device communication,'' \emph{IEEE Wireless
  Commun.}, vol.~21, no.~3, pp. 136--144, June 2014.

\bibitem{SSFD}
X.~Lin, J.~Andrews, and A.~Ghosh, ``Spectrum sharing for device-to-device
  communication in cellular networks,'' \emph{IEEE Trans. Wireless Commun.},
  vol.~13, no.~12, pp. 6727--6740, Dec. 2014.

\bibitem{AMMS}
H.~ElSawy, E.~Hossain, and M.-S. Alouini, ``Analytical modeling of mode
  selection and power control for underlay {D2D} communication in cellular
  networks,'' \emph{IEEE Trans. Commun.}, vol.~62, no.~11, pp. 4147--4161, Nov
  2014.

\bibitem{ROID2D}
Q.~Ye, M.~Al-Shalash, C.~Caramanis, and J.~Andrews, ``Resource optimization in
  device-to-device cellular systems using time-frequency hopping,'' \emph{IEEE
  Trans. Wireless Commun.}, vol.~13, no.~10, pp. 5467--5480, Oct 2014.

\bibitem{ERAF_Song}
C.~Xu, L.~Song, Z.~Han, Q.~Zhao, X.~Wang, X.~Cheng, and B.~Jiao, ``Efficiency
  resource allocation for device-to-device underlay communication systems: A
  reverse iterative combinatorial auction based approach,'' \emph{IEEE J. Sel.
  Areas Commun.}, vol.~31, no.~9, pp. 348--358, Sept. 2013.

\bibitem{CGFRA}
Y.~Li, D.~Jin, J.~Yuan, and Z.~Han, ``{Coalitional Games for Resource
  Allocation in the Device-to-Device Uplink Underlaying Cellular Networks},''
  \emph{IEEE Trans. Wireless Commun.}, vol.~13, no.~7, pp. 3965--3977, July
  2014.

\bibitem{OTCR}
M.~Sheng, J.~Liu, X.~Wang, Y.~Zhang, H.~Sun, and J.~Li, ``On transmission
  capacity region of {D2D} integrated cellular networks with interference
  management,'' \emph{accepted by IEEE Trans. Commun.}, 2015.

\bibitem{SGAI}
W.~K. D.~Stoyan and J.~Mecke, \emph{{Stochastic geometry and its
  applications}}.\hskip 1em plus 0.5em minus 0.4em\relax Chichester: Wiley,
  1995.

\bibitem{LTUL}
S.~Sesia, I.~Toufik, and M.~Baker, \emph{{LTE- The UMTS Long Term Evolution:
  From Theory to Practice}}.\hskip 1em plus 0.5em minus 0.4em\relax New Jersey,
  NJ, USA: John Wiley \& Sons, 2011.

\bibitem{DUDC}
Z.~Shen, A.~Khoryaev, E.~Eriksson, and X.~Pan, ``Dynamic uplink-downlink
  configuration and interference management in {TD-LTE},'' \emph{IEEE Commun.
  Mag.}, vol.~50, no.~11, pp. 51--59, Nov. 2012.

\bibitem{DUDO}
M.~S. ElBamby, M.~Bennis, W.~Saad, and M.~Latva-aho, ``{Dynamic Uplink-Downlink
  Optimization in TDD-based Small Cell Networks},'' in \emph{Proc. IEEE INFOCOM
  Workshops}, Toronto, Canada, Apr. 27 - May 2, 2014, pp. 712--717.

\bibitem{DTAF}
J.~Li, S.~Farahvash, M.~Kavehrad, and R.~Valenzuela, ``Dynamic {TDD} and fixed
  cellular networks,'' \emph{IEEE Commun. Lett.}, vol.~4, no.~7, pp. 218--220,
  Jul. 2000.

\bibitem{DTSI}
B.~Yu, S.~Mukherjee, H.~Ishii, and L.~Yang, ``{Dynamic TDD support in the LTE-B
  enhanced Local Area architecture},'' in \emph{IEEE GLOBECOM Workshops},
  Anaheim, America, Dec. 3-7, 2012, pp. 585--591.

\bibitem{CHDD}
Y.~S. Soh, T.~Q.~S. Quek, M.~Kountouris, and G.~Caire, ``Cognitive hybrid
  division duplex for two-tier femtocell networks,'' \emph{IEEE Trans. Wireless
  Commun.}, vol.~12, no.~10, pp. 4852--4865, Oct. 2013.

\bibitem{PCIT}
V.~Chandrasekhar, J.~G. Andrews, T.~Muharemovict, Z.~Shen, and A.~Gatherer,
  ``{Power control in two-tier femtocell networks},'' \emph{IEEE Trans.
  Wireless Commun.}, vol.~8, no.~8, pp. 4316--4328, Aug. 2009.

\bibitem{SICIH}
M.~Wildemeersch, T.~Q.~S. Quek, M.~Kountouris, A.~Rabbachin, and C.~Slump,
  ``Successive interference cancellation in heterogeneous networks,''
  \emph{IEEE Trans. Commun.}, vol.~62, no.~12, pp. 4440--4453, Dec. 2014.

\bibitem{SAIT}
V.~Chandrasekhar and J.~G. Andrews, ``{Spectrum allocation in tiered cellular
  networks},'' \emph{IEEE Trans. Wireless Commun.}, vol.~57, no.~10, pp.
  3059--3068, Oct. 2009.

\bibitem{ASGM}
T.~V. Nguyen and F.~Baccelli, ``A stochastic geometry model for cognitive radio
  networks,'' \emph{The Computer Journal}, vol.~55, no.~5, pp. 534 -- 552, Jul.
  2012.

\bibitem{MIIH}
M.~Haenggi, ``Mean interference in hard-core wireless networks,'' \emph{IEEE
  Commun. Lett.}, vol.~15, no.~8, pp. 792--794, Aug. 2011.

\bibitem{AMHC}
H.~ElSawy and E.~Hossain, ``{A Modified Hard Core Point Process for Analysis of
  Random CSMA Wireless Networks in General Fading Environments},'' \emph{IEEE
  Trans. Commun.}, vol.~61, no.~4, pp. 1520--1534, Apr. 2013.

\bibitem{IAOI}
C.~han Lee and M.~Haenggi, ``Interference and outage in poisson cognitive
  networks,'' \emph{IEEE Trans. Wireless Commun.}, vol.~11, no.~4, pp.
  1392--1401, Apr. 2012.

\bibitem{HCNW}
H.-S. Jo, Y.~J. Sang, P.~Xia, and J.~G. Andrews, ``Heterogeneous cellular
  networks with flexible cell association: A comprehensive downlink {SINR}
  analysis,'' \emph{IEEE Trans. Wireless Commun.}, vol.~11, no.~10, pp.
  3484--3495, Oct. 2012.

\bibitem{AMUC}
T.~Novlan, H.~Dhillon, and J.~Andrews, ``{Analytical Modeling of Uplink
  Cellular Networks},'' \emph{IEEE Trans. Wireless Commun.}, vol.~12, no.~6,
  pp. 2669--2679, June 2013.

\bibitem{ATAT}
J.~G. Andrews, F.~Baccelli, and R.~Ganti, ``{A Tractable Approach to Coverage
  and Rate in Cellular Networks},'' \emph{IEEE Trans. Commun.}, vol.~59,
  no.~11, pp. 3122--3134, Nov. 2011.

\bibitem{OIHN}
S.~Singh, H.~S. Dhillon, and J.~G. Andrews, ``Offloading in heterogeneous
  networks: Modeling, analysis, and design insights,'' \emph{IEEE Trans.
  Wireless Commun.}, vol.~12, no.~5, pp. 2484--2497, May 2013.

\bibitem{DCAB}
S.~M. Yu and S.-L. Kim, ``Downlink capacity and base station density in
  cellular networks,'' in \emph{Proc. IEEE WiOpt}, Tsukuba Science City, Japan,
  May 13-17, 2013, pp. 119--124.

\bibitem{MAKT}
H.~Dhillon, R.~Ganti, F.~Baccelli, and J.~Andrews, ``Modeling and analysis of
  {K-Tier} downlink heterogeneous cellular networks,'' \emph{IEEE J. Sel. Areas
  Commun.}, vol.~30, no.~3, pp. 550--560, 2012.

\end{thebibliography}

\end{document}